\documentclass[graybox]{svmult}

\usepackage{amssymb}
\usepackage{amsmath}
\usepackage{slashed}
\usepackage{multirow}
\usepackage[table]{xcolor}

\usepackage{mathptmx}       
\usepackage{helvet}         
\usepackage{courier}        
\usepackage{type1cm}        
\usepackage{makeidx}         
\usepackage{graphicx}        
\usepackage{multicol}        
\usepackage[bottom]{footmisc}

\makeindex             

\newcommand{\beqn}{\begin{eqnarray}}
\newcommand{\eeqn}{\end{eqnarray}}
\newcommand{\beqs}{\begin{subequations}}
\newcommand{\eeqs}{\end{subequations}}
\newcommand{\eq}[1]{(\ref{#1})}
\newcommand{\cD}{{\mathfrak D}}

\newcommand{\ext}{{\mathrm{ext}}}
\newcommand{\cL}{{\cal L}}
\newcommand{\cl}{{\ell}}
\newcommand{\cO}{{\cal O}}

\newcommand{\cM}{{\cal M}}
\newcommand{\cZ}{{\cal Z}}
\newcommand{\GN}{{\mathrm{GN}}}
\newcommand{\NJL}{{\mathrm{NJL}}}

\newcommand{\oeps}{\overline{\epsilon}}

\newcommand{\dd}{\!{\mathrm d}}

\renewcommand{\matrix}[4]{\left(\begin{array}{cc} #1 & #2 \\ #3 & #4\end{array}\right)}

\def\bbbone{{\mathchoice {\rm 1\mskip-4mu l} {\rm 1\mskip-4mu l} {\rm 1\mskip-4.5mu l} {\rm 1\mskip-5mu l}}}
\newcommand{\tr}{{\mathrm{tr}}}
\newcommand{\Z}{{Z \!\!\! Z}}
\def\bbbone{{\mathchoice {\rm 1\mskip-4mu l} {\rm 1\mskip-4mu l}
{\rm 1\mskip-4.5mu l} {\rm 1\mskip-5mu l}}}

\newcommand{\nhline}{\hline\noalign{\smallskip}}
\newcommand{\mline}{\noalign{\smallskip}\hline\noalign{\smallskip}}
\newcommand{\mmline}{\noalign{\smallskip}\svhline\noalign{\smallskip}}

\begin{document}

\title*{Electromagnetic superconductivity of vacuum induced by strong magnetic field}
\author{M. N. Chernodub}
\institute{M. N. Chernodub \at 
CNRS, Laboratoire de Math\'ematiques et Physique Th\'eorique, Universit\'e Fran\c{c}ois-Rabelais, F\'ed\'eration Denis Poisson, Parc de Grandmont, 37200 Tours, France;
Department of Physics and Astronomy, University of Gent, Krijgslaan 281, S9, 9000 Gent, Belgium; on leave from ITEP, B. Cheremushkinskaya 25, 117218 Moscow, Russia, 
\email{maxim.chernodub@lmpt.univ-tours.fr}.
This work was supported by Grant No. ANR-10-JCJC-0408 HYPERMAG.}
%
%
\maketitle

\vskip -25mm 
\abstract{The quantum vacuum may become an electromagnetic superconductor in the presence of a strong external magnetic field of the order of $10^{16}$ Tesla. The magnetic field of the required strength (and even stronger) is expected to be generated for a short time in ultraperipheral collisions of heavy ions at the Large Hadron Collider. The superconducting properties of the new phase appear as a result of a magnetic--field--assisted condensation of quark-antiquark pairs with quantum numbers of electrically charged $\rho^\pm$ mesons. We discuss similarities and differences between the suggested superconducting state of the quantum vacuum, a conventional superconductivity and the Schwinger pair creation. We argue qualitatively and quantitatively why the superconducting state should be a natural ground state of the vacuum at the sufficiently strong magnetic field. We demonstrate the existence of the superconducting phase using both the Nambu--Jona-Lasinio model and an effective bosonic model based on the vector meson dominance (the $\rho$--meson electrodynamics).  We discuss various properties of the new phase such as absence of the Meissner effect, anisotropy of superconductivity, spatial inhomogeneity of ground state, emergence of a neutral superfluid component in the ground state and presence of new topological vortices in the quark-antiquark condensates.}

\abstract*{The quantum vacuum may become an electromagnetic superconductor in the presence of a strong external magnetic field of the order of $10^{16}$ Tesla. The magnetic field of the required strength (and even stronger) is expected to be generated for a short time in ultraperipheral collisions of heavy ions at the Large Hadron Collider. The superconducting properties of the new phase appear as a result of a magnetic--field--assisted condensation of quark-antiquark pairs with quantum numbers of electrically charged $\rho^\pm$ mesons. We discuss similarities and differences between the suggested superconducting state of the quantum vacuum, a conventional superconductivity and the Schwinger pair creation. We argue qualitatively and quantitatively why the superconducting state should be a natural ground state of the vacuum at the sufficiently strong magnetic field. We demonstrate the existence of the superconducting phase using both the Nambu--Jona-Lasinio model and an effective bosonic model based on the vector meson dominance (the $\rho$--meson electrodynamics).  We discuss various properties of the new phase such as absence of the Meissner effect, anisotropy of superconductivity, spatial inhomogeneity of ground state, emergence of a neutral superfluid component in the ground state and presence of new topological vortices in the quark-antiquark condensates.}

\vskip 10mm 
{\small{
\noindent To appear in Lect. Notes Phys. "Strongly interacting matter in magnetic
fields" (Springer), edited by D. Kharzeev, K. Landsteiner, A. Schmitt,
H.-U. Yee }}

\vskip -5mm

\section{Introduction}

Quantum Chromodynamics (QCD) exhibits many remarkable properties in the presence of a very strong magnetic field. The external magnetic field affects dynamics of quarks because the quarks are electrically charged particles. As a result, the magnetic field enhances the chiral symmetry breaking by increasing the value of the (quark) chiral condensate~\cite{ref:Igor:review}. The change in the dynamics of quarks is also felt by the gluon sector of QCD because the quarks are coupled to the gluons. Therefore, the magnetic field may affect the whole strongly interacting sector and influence very intrinsic properties of QCD such as, for example, the confinement of color~\cite{ref:Marco:review,ref:Eduardo:review}. 
  
In order to make a noticeable influence on the strongly interacting sector, the strength of the magnetic field should be of the order of a typical QCD mass scale, $e B \sim m_{\pi}^{2}$, where $m_{\pi} \approx 140\, \mathrm{MeV}$ is the pion mass. The corresponding magnetic field strength, $B \sim 3 \times 10^{14}\, \mathrm{T}$, is enormous from a human perspective ($1 \mathrm{T} \equiv 10^{4}\, \mathrm{G}$). However, such strong magnetic field can be achieved in noncentral heavy-ion collisions at Relativistic Heavy-Ion Collider (RHIC)~\cite{Skokov:2009qp}. At higher energies of the Large Hadron Collider (LHC), noncentral heavy-ion collisions may generate even higher magnetic field of $e B \sim 15 m_{\pi}^{2}$ ($B \sim 5 \times 10^{15}\, \mathrm{T}$)~\cite{Skokov:2009qp}.  And in ultraperipheral collisions -- when two nuclei pass near each other without a real collision -- the magnetic field strength may even reach $e B \sim (60 \dots 100) m_{\pi}^{2}$ or, in conventional units, $B \sim (2 \dots 3) \times 10^{16}\, \mathrm{T}$, Ref.~\cite{Deng:2012pc}.

Despite the magnetic field is generated in a heavy-ion collision for a very short time, it may have observable consequences. In noncentral collisions, the magnetic field is generated together with a hot expanding fireball of quark-gluon plasma. Topological QCD transitions may lead to a chiral imbalance of the plasma, and, in turn, the chirally--imbalanced matter may produce an observable electric current along the axis of the magnetic field~\cite{Vilenkin:1980fu} driven by the chiral magnetic effect~\cite{Fukushima:2008xe}.

In a finite--density (quark) matter the magnetic catalysis~\cite{ref:Igor:review} may be reversed~\cite{ref:Andreas:review} and the phase diagram may be modified substantially~\cite{ref:Andreas:review,ref:Johanna:review}. In the absence of matter (i.e., in the vacuum), the external magnetic field affects the finite--temperature phase structure of the theory by shifting the critical temperatures and altering the strength of the confinement--deconfinement and chiral transitions~\cite{ref:Marco:review,ref:Eduardo:review}. 

The vacuum may also spontaneously become an electromagnetic superconductor if the magnetic field strength exceeds the following critical value~\cite{Chernodub:2010qx,Chernodub:2011mc}:
\beqn
B_c \simeq 10^{16} \, {\mathrm{Tesla}}
\qquad  {\mathrm{or}} \qquad
e B_c \simeq 0.6\,\mbox{GeV}^2\,.
\quad
\label{eq:Bc}
\eeqn 
The counterintuitive superconductivity of, basically, empty space, should always be accompanied by a superfluid component~\cite{Chernodub:2011tv,ref:Jos}. We discuss these effects below.

In Section~\ref{sec:comparison} we describe the mechanism and the basic features of the vacuum superconductivity in a very qualitative way. We compare in details the vacuum superconductivity with an ordinary superconductivity. We also highlight certain similarities of this exotic vacuum phase with a magnetic-field-assisted ``reentrant superconductivity''  in condensed matter and the electric-field-induced Schwinger electron-positron pair production in the vacuum of Quantum Electrodynamics. 

In Section~\ref{sec:quantitative} the emergence of the superconducting phase is explicitly demonstrated both in a bosonic $\rho$--meson electrodynamics~\cite{Djukanovic:2005ag}
and in an extended Nambu--Jona-Lasinio model~\cite{Ebert:1985kz}. Various properties of the superconducting state are summarized in the last Section.

\clearpage

\section{Conventional superconductivity, vacuum superconductivity and Schwinger pair creation: differences and similarities}
\label{sec:comparison}

\subsection{Conventional superconductivity via formation of Cooper pairs}

Before going into details of the magnetic--field--induced vacuum superconductivity let us discuss certain very basic qualitative features of the conventional superconductivity. Why certain compounds are superconductors?

In a simplified picture\footnote{A comprehensive introduction to superconductivity can be found in books~\cite{ref:LL:9,ref:Abrikosov,ref:Tinkham}.}, electrons in a metal can be considered as (almost free) negatively charged particles which move through a periodically structured background of a lattice made of positively charged ions. The individual electrons scatter inelastically off the ions leading to a dissipation of an electric current and, consequently, to emergence of a nonvanishing electrical resistance in the metal. 

\begin{figure}[b]
\begin{center}
\begin{tabular}{lr}
\includegraphics[scale=.35, angle = 90]{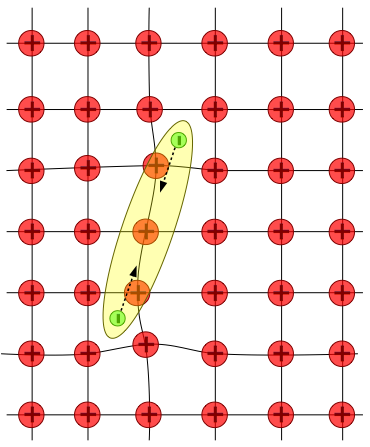} &
\hskip 5mm \includegraphics[scale=.15]{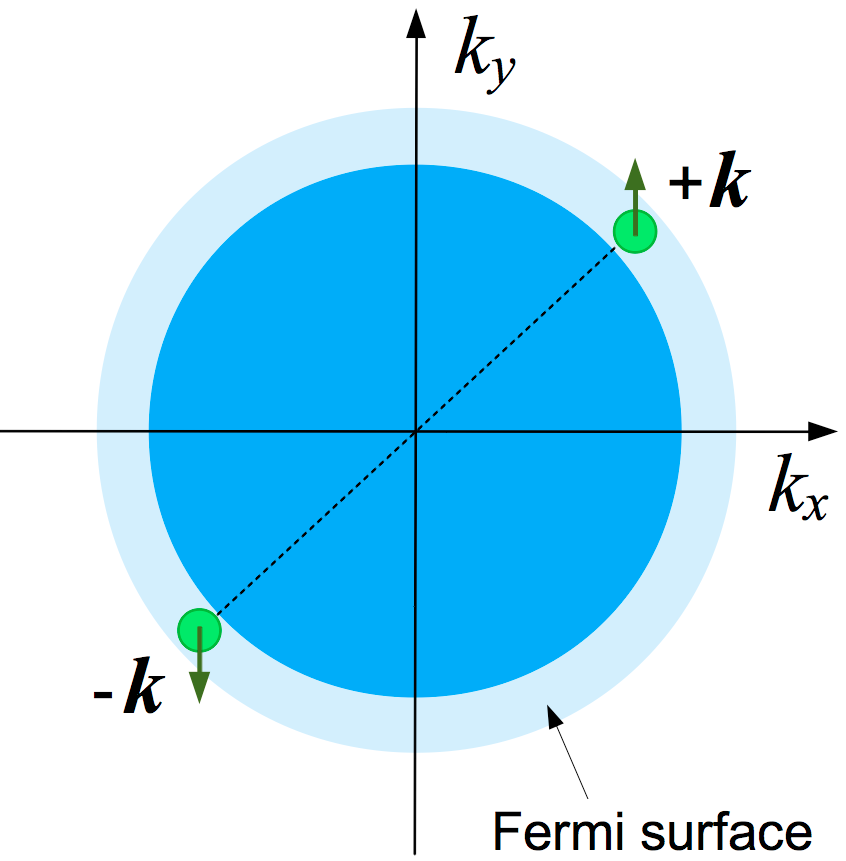} 
\end{tabular}
\end{center}
\caption{(left) Formation of the Cooper pair (the yellowish oval) of electrons (the small green circles) in an ionic lattice (the large red circles) due to phonon interaction.
(right) Two interacting electrons (the small green circles) and the Fermi sphere (the large blue circle) in the momentum space. The electrons in the Cooper pair have mutually opposite spins (the green arrows) and momenta.
}
\label{fig:Cooper:lattice}     
\label{fig:Fermi:sphere}     
\end{figure}

As an electron moves through the ionic lattice it attracts neighboring ions via a Coulomb interaction. The attraction leads to a  local deformation of the ionic lattice, and, simultaneously, to an excess of the positive electric charge in a vicinity of the electron. The excess of the positive charge, in turn, attracts another electron nearby, so that in a background of the positively charged ion lattice the like-charged electrons may experience a mutual attractive force, Fig.~\ref{fig:Cooper:lattice}(left). The deformation of the ionic lattice can be viewed as a superposition of collective excitations of the ion lattice (phonons), so that the process of the electron-electron interaction via lattice deformations can be described by a phonon exchange.

The attractive force between the electrons may, in principle, lead to formation of electron--electron bound states. However, this attraction is extremely weak and therefore thermal fluctuations at room temperature easily destroy the two--electron bound states. On the other hand, at low temperature the attractive interaction between the electrons prevails the thermal disorder and, consequently, the bound states may indeed be formed. These bound state are called the Cooper pairs. The electrons in the Cooper pairs have mutually opposite spins and opposite momenta thus making the Cooper pair a (composite) spin-zero bosonic state. Bosons have a tendency to condense at low temperature so that the Cooper pair condensate may emerge.

In the condensed state all Cooper pairs behave as one collective entity. The Cooper--pair condensate can move frictionlessly through the ion lattice similarly to a motion of a superfluid. The motion without dissipation is guaranteed due to an energy gap, which separates the energy of the condensed ground state and a next, excited state. Since the intermediate states are absent, at certain conditions (low enough temperature, weak enough electric current, etc) the dissipative scattering of the Cooper pairs off the ions becomes kinematically impossible so that the motion of the Cooper pair condensate proceeds without dissipation.  Since the Cooper pairs are electrically charged states, the condensation of the Cooper pairs turns the material into a superconducting  state.

The formation of the Cooper pairs is facilitated by the fact that at low temperature the dynamics of the electrons becomes effectively one dimensional while in one spatial dimension even a very weak attraction between two particles should always lead to formation of a bound state (in the condensed matter context this property is known as the Cooper theorem). The effective dimensional reduction of the electron dynamics from three spatial dimensions to one spatial dimension is possible because at low temperature the interaction between the electrons occurs if and only if the electrons are sufficiently close to the Fermi surface. One component of the momentum at the Fermi surface counts the degeneracy of the electron states while the other component is a dynamical degree of freedom. The Cooper pair is formed by two electrons with mutually opposite momenta and mutually opposite spins, Figure~\ref{fig:Fermi:sphere}(right).

Summarizing, in order to exhibit the conventional superconductivity a system should satisfy the following basic requirements:
\begin{itemize}
\item[A)]\ electric charge carriers should be present in the system (otherwise the system cannot support the electric current);
\label{page:condition:A}
\item[B)]\ dynamics of the electric charge carriers should effectively be one--dimensional (otherwise the Cooper pairs cannot be formed);
\label{page:condition:B}
\item[C)]\ the like--charged carriers should experience mutual (pairwise) attraction (otherwise the Cooper pairs cannot be formed).
\label{page:condition:C}
\end{itemize}
\label{page:requirements}

Surprisingly, the same requirements are satisfied by the quantum fluctuations of the vacuum in a background of a sufficiently strong magnetic field. In the next section we compare basic features of conventional and ``vacuum'' superconductors. 

\subsection{Vacuum superconductivity}

\subsubsection{Condition A: presence of electric charges}
\label{ref:environment}

In order for the vacuum to behave as an electromagnetic superconductor, one needs, at least, the presence of electrically charged particles ({\bf condition A} on page \pageref{page:requirements}). From a first sight, it is impossible to satisfy this requirement because under the usual conditions the vacuum is characterized by the absence of the free electric charges. Nevertheless, the quantum vacuum may be considered as an excellent ``reservoir'' of various particles including the electrically charged ones. Moreover, under certain external conditions the virtual particles may become real. 

This ``virtual-to-real'' scenario does not sound unlikely.  For example, there are at least two well-known cases of external conditions when a vacuum becomes an electrically conducting media: the vacuum may conduct electricity if it is either subjected under a strong electric field or if it is sufficiently hot.

The first example of the ``virtual-to-real'' transition is the Schwinger effect in Quantum Electrodynamics (QED): a sufficiently strong external electric field generates electron-positron pairs out of the vacuum~\cite{ref:Schwinger}. The created positrons and electrons move in opposite directions thus creating an electric current\footnote{We briefly discuss an analogy between the magnetic-field-induced vacuum superconductivity and the Schwinger effect in Section~\ref{sec:Schwinger}, page~\pageref{sec:Schwinger}.}. The critical strength of the electric field required for this process is $E_{c} = m^{2}_{e}/e\approx 10^{18}$\,V/m.

The second ``virtual-to-real'' example is a simple thermal ionization of electron--positron pairs: the vacuum turns into an electron--positron plasma at temperatures $T \sim 0.1 \, T^{\mathrm{QED}}$ where $T^{\mathrm{QED}} \approx 2 m_e \approx 1\,\mbox{MeV} \approx 10^{10}\,\mathrm{K}$ is a typical QED temperature.

\begin{table}
\caption{Conventional superconductivity vs. vacuum superconductivity: very general features (from Ref.~\cite{Chernodub:2012gr}).}
\label{tbl:environment}
\begin{tabular}{p{2.8cm}p{4.3cm}p{3.7cm}}
\mline
Property & Conventional superconductivity \qquad \qquad & Vacuum superconductivity \\[1mm]
\mmline
Environment &  a material (metal, alloy etc) &   vacuum (empty space)\\[1mm]
\nhline
Reservoir of carriers &  real particles & virtual particles \\[1mm]
\nhline
Normal state \qquad \qquad &  a conductor &  an insulator\\[1mm]
\nhline
Basic carriers  of & \multirow{2}{*}{electrons ($e$)} & light quarks ($u$, $d$) and \\[1mm]
electric charge &  & light antiquarks  ($\bar u$, $\bar d$) \\[1mm]
\nhline
Electric charges    &  \multirow{2}{*}{ $q_e = - e$ \quad ($e \equiv |e|$)} &  $q_u = + 2 e/3$,\ \ $q_d = - e/3$  \\[1mm]
of basic carries     &  &  $q_{\bar u} = - 2 e/3$,\ \  $q_{\bar d} = + e/3$  \\[1mm]
\nhline
\end{tabular}
\end{table}

Thus, the quantum vacuum may be turned into a conductor if it is subjected to sufficiently strong electric field ($E \sim 10^{18}$\,V/m) or to sufficiently high temperature ($T \sim 10^{9}$\,K). Below we show that sufficiently strong magnetic field ($B \sim 10^{16}$\,T) may turn the vacuum into a superconducting state. The magnetic--field--induced vacuum superconductivity works at the QCD scale: the key role here is played by virtual quarks and antiquarks which have fractional electric charges. As we discuss below, the strong magnetic field catalyses the formation of the electrically charged condensates made of quarks and antiquarks. Very general features of a conventional superconductor and the magnetic--field--induced vacuum superconductivity are summarized in Table~\ref{tbl:environment}.

\subsubsection{Conditions B and C: formation of superconducting carriers}

In order for the superconducting carriers to be formed, the fermion dynamics should be reduced from three spatial dimensions to one spatial dimension ({\bf condition B} of Section~\ref{ref:environment}, page~\pageref{page:condition:B}). In conventional superconductivity the dimensional reduction proceeds via formation of the Fermi surface at sufficiently low temperatures. This mechanism cannot work in our case because the Fermi surface, obviously, does not exist in the vacuum due the very absence of matter. However, the dimensional reduction may be achieved with the help of a magnetic field background since electrically charged particles with low-energy can move only along the axis of the magnetic field. This effect leads to the required dimensional reduction of the charge's dynamics from three to one spatial dimensions.

The described dimensional reduction effect in the background of the external magnetic field works for all electrically charged elementary particles, including electrons, positrons, quarks, antiquarks etc. However, the superconducting bound state may only be formed from a particular combination of these particles which should satisfy the following conditions:
\begin{itemize}
\item[(i)] the superconducting bound state should be a boson;
\item[(ii)] ~the bound state should be electrically charged;
\item[(iii)] ~~the interaction between the constituents of the bound state should be attractive.
\end{itemize}

Condition (i) implies that the superconducting bound state should contain even number of constituents because the known carriers of the electric charge are fermions (quarks, electron and positron, etc). Below we consider simplest, two-fermion states.

Condition (ii) implies that the bound state cannot be composed of a particle and its antiparticle. In combination with condition (iii) it means that the vacuum superconductivity cannot -- unlike the Schwinger's pair creation -- emerge in the pure QED vacuum sector which describes electrons, positrons and photons. Indeed, the electron--electron interaction is mediated by a repulsive photon exchange so that condition (iii) is not satisfied. On the other hand, the interaction between electron and positron is attractive, but the electron--positron bound state is electrically neutral so that in this case condition (ii) is not satisfied. Thus, the superconductivity cannot emerge in the pure QED.

Therefore, the candidates for the superconducting charged bound states should be outside of the purely electrodynamics sector. Below we concentrate on the next (in terms of energy scale), strongly interacting sector which describes the dynamics of quarks and gluons.

The QCD sector of the vacuum contains the gluon particle which is a carrier of the strong force. From our perspective the gluon is an analogue of the phonon of conversional superconductivity because it is the gluon may which may provide an attractive interaction between quarks and antiquarks regardless of their electric charges. In particular, the gluon may bind a quark and an anti-quark into an electrically charged meson. The attractive nature of the gluon interaction allows us to satisfy {\bf condition~C} of superconductivity on page~\pageref{page:condition:C}. 

Thus, the suggested mechanism of the vacuum superconductivity may indeed work at the interface of the QED and QCD sectors. The simplest example of the superconducting carrier may be given by a bound state of a $u$ quark with the electric charge $q_u = + 2 e/3$ and a $\bar d$ antiquark with the electric charge $q_{\bar d} \equiv - q_d = + e/3$. The attractive nature of the gluon--mediated interaction between the quark and antiquark of different flavors is only possible if these constituents reside in a triplet state, so that the $u \bar d$ bound state should be a spin-1 state (the $\rho$ meson). 

Therefore, the vacuum analogue of the Cooper pair are the charged $\rho^{\pm}$ meson states. And in next sections we show that the $\rho$--meson condensates do indeed appear in the vacuum in the presence of the strong magnetic field, and we argue that the emergent state is indeed an electromagnetic superconductor.

\begin{table}
\caption{Conventional superconductivity vs. vacuum superconductivity: superconducting carriers}
\label{tbl:carriers}
\begin{tabular}{p{3.0cm}p{4.3cm}p{3.9cm}}
\mline
Property of carrier & Conventional superconductivity \qquad \qquad & Vacuum superconductivity \\[1mm]
\mmline
Type  & Cooper pair & $\rho$--meson excitations, $\rho^+$ and $\rho^-$ \\[1mm]
\nhline
\multirow{2}{*}{Composition}   & \multirow{2}{*}{electron-electron state ($ee$)} & quark-antiquark states\\[1mm]
&   & ($\rho^+ = u\bar d$ and $\rho^- = d\bar u$) \\[1mm]
\nhline
Electric charge  & $- 2 e$ & $+ e$ and $-e$, respectively\\[1mm]
\nhline
Spin  & typically spin-zero state (scalar) & one-one state (vector) \\[1mm]
\nhline
                                                            & \multicolumn{2}{l}{1) reduction of dynamics of basic electric charges}\\[1mm]
   The carriers are                  & \multicolumn{2}{l}{\phantom{1) }from three spatial dimensions to one dimension, $3d \to 1d$}	\\[1mm]
\cline{2-3}\\[-3mm]   
   formed due to                               & 2) attraction force between                             &  2) attraction force between \\[-3mm]
                                                            & \phantom{2) } two electrons &  \phantom{2) }a quark and an antiquark \\[1mm]
\nhline
\multirow{3}{*}{
\begin{tabular}{l}
\hskip 0mm 1) a reason for the \\[1mm] 
\hskip 0mm reduction $3d \to 1d$
\end{tabular}} 
 & at very low temperatures &  in strong magnetic field the\\[1mm]
 &  electrons interact with each  &  motion of electrically charged \\[1mm]
& other near the Fermi surface  & particles is one dimensional \\[1mm]
\cline{2-3}\\[-3mm]
\hskip 1mm 2) attraction is due to & phonons (lattice vibrations) & gluons (strong force, QCD)\\[1mm]
\hline\\[-2mm]
Isotropy of & \multirow{3}{*}{\begin{tabular}{c} Yes: superconducting \\[1mm] \ in all spatial directions \end{tabular}} & No: superconducting along \\[1mm]
superconducting &  & the axis of the magnetic field,\\[1mm]
properties &  & insulator in other directions\\[1mm]
\nhline
\end{tabular}
\end{table}

Notice that the formation of the bound state is facilitated by the dimensional reduction of the quark's dynamics in the background of the magnetic field ({\bf condition~B}). The dimensional reduction implies automatically a strong anisotropy of the suggested superconductivity since the electric charges (the quarks $u$ and $d$ and their antiquarks) may move only along the axis of the magnetic field. As a result, the superconducting charge carriers (the $\rho$ mesons in our case) may also flow along the axis of the magnetic field only. Thus, the vacuum exhibit a superconducting property in the longitudinal direction (along the magnetic field) while in the two transverse directions the superconductivity of the vacuum should be absent.

Note that due to the anisotropic superconducting properties the vacuum in the strong magnetic field acquires a very unusual optical property: the vacuum becomes as (hyperbolic) metamaterial which behaves as diffractionless ``perfect lenses'' \cite{Smolyaninov:2011wc}. 

In Table~\ref{tbl:carriers} we compare of certain basic features of the superconducting carriers in a conventional (low-temperature) superconductivity and in the vacuum (high-magnetic-field) superconductivity.

\subsubsection{Counterintuitive coexistence of magnetic field and superconductivity due to strong anisotropy of magnetic-field-induced superconductivity}
\label{sec:coexistence}

So far we have ignored a well-known property of all known superconductors: 
\begin{itemize}
\item Magnetic field (regardless of its strength) and conventional superconductivity\footnote{Except for the unconventional reentrant superconductivity (Section~\ref{sec:reentrant}, page~\pageref{sec:reentrant}).} (regardless of its mechanism) cannot coexist with each other!
\end{itemize}
Thus, we can ask ourselves: why do we believe that the superconducting phase of the vacuum can exist in (and, moreover, be induced by) the strong magnetic field? In fact, this single question contains two puzzles (Table~\ref{tbl:magnetic}):
\begin{itemize}
\item Why the Meissner effect is absent in the superconducting phase of vacuum? 
\item Why strong magnetic field does not destroy the superconductivity of vacuum?
\end{itemize}

A short ``technical'' answer to these questions is that in the background of the magnetic field the superconducting state of the vacuum has lower energy compared to the energy of the normal (insulator) state (Section~\ref{sec:energetic}, page~\pageref{sec:energetic}). A physical argument is that the strong magnetic field may coexist with the vacuum superconductivity
because the latter is highly anisotropic. Let us consider this point in detail.

\begin{table}
\caption{A comparison of the effects of magnetic field and thermal effects on conventional superconductivities and electromagnetic superconductivity of vacuum (from Ref.~\cite{Chernodub:2012gr}).}
\label{tbl:magnetic}
\begin{tabular}{p{32mm}p{45mm}p{35mm}}
\mline
Property & Conventional superconductivity \qquad \qquad & Vacuum superconductivity \\[1mm]
\mmline
Magnetic field & destroys superconductivity & induces superconductivity \\[1mm]
\mline
The Meissner effect & present & absent \\[1mm] 
\nhline
Thermal fluctuations & destroy superconductivity & destroy superconductivity \\[1mm] 
\mline
\end{tabular}
\end{table}

Qualitative arguments against the Meissner effect in the vacuum superconductor are as follows. The Meissner effect is a screening of weak external magnetic field by a superconducting state so that a magnetic field cannot penetrate deeply into a superconductor. Qualitatively, the Meissner effect is caused by superconducting currents which are induced by the external magnetic field in the bulk of a superconductor. The circulation of these currents in the transversal (with respect to the magnetic field axis) plane generates a backreacting magnetic field, which screens the external magnetic field in the bulk of the superconducting material. The backreacting currents are geometrically large, so that the corresponding magnetic length (i.e., the radius of the lowest Landau level), $1/\sqrt{|e B|}$, is much larger than the correlation length $\xi$ of the superconductor. Since the vacuum superconductivity is realized only along the axis of the magnetic field, the large transversal currents are absent and the Meissner effect cannot be realized. 

 If the axis of the external magnetic field is oriented along the normal to a boundary of an ordinary superconductor, then the backreacting magnetic field squeezes the external magnetic field into thin Abrikosov vortices\footnote{Since the magnetic flux coming through the superconductor's boundary is a conserved quantity, the superconductor expels it from the superconductor's bulk into thin vortexlike structures.} which form a sparse vortex lattice in a background of weak magnetic field.  As we show below (Section~\ref{sec:vortices}, page~\pageref{sec:vortices}), the superconducting ground state of the vacuum is a dense lattice of Abrikosov-type vortices for which the magnetic length is of the order of (or even smaller than) the correlating length. This is a quantum regime of the Abrikosov lattice, in which the geometrically short transverse currents in the cannot screen geometrically large external magnetic field. In this case the physical situation is similar to a ``reentrant superconductivity''of  extreme type-II superconductor in a high magnetic field~\cite{ref:Tesanovic} (Section~\ref{sec:reentrant}, page~\pageref{sec:reentrant}).

If the axis of the weak external magnetic field is oriented tangentially to a boundary of an ordinary superconductor then the external field is usually expelled from the superconductor's bulk without formation of the Abrikosov vortices. In our case, the superconducting ground state is created by a strong magnetic field, and therefore the very imposition of a weak tangential magnetic field is logically impossible from very simple geometrical reasons: a sole result of the superposition of the weak ``testing'' magnetic field onto the strong ``creating'' magnetic field is a slight turn of the stronger field. In other words, the weak testing field should slightly reorientate the anisotropy axis of the superconductor without destroying it.

The ordinary superconductivity is destroyed by sufficiently strong magnetic field. Qualitatively, one can understand this effect as follows: in strong enough external magnetic field the (positive) excess in energy of the induced transverse superconducting currents prevails the (negative) condensation energy of the superconducting carriers. As a result, at certain critical field the conventional superconductivity becomes energetically unfavorable and the material turns from the superconducting state back to the normal (nonsuperconducting) state. On the contrary, in the vacuum superconductivity the large superconducting currents are absent due to strong anisotropy of the superconducting currents, so that the mentioned argument should not work. Moreover, as we   discuss below, the energy of the short transverse currents is diminished as the magnetic field becomes stronger.

Thus, the electromagnetic superconductivity of the vacuum coexists with high magnetic field due to the anisotropy of the magnetic-field-induced superconducting properties. It is the anisotropy which makes the vacuum superconductivity to be different from the conventional one.

\subsubsection{Magnetic-field-induced vacuum superconductivity: temperature effects}

The common key element of the ordinary and vacuum superconductivities is the dimensional reduction of the dynamics of the charge carriers (condition B on page~\pageref{page:requirements}). Thermal fluctuations should destroy this property regardless of the mechanism of the dimensional reduction. In ordinary superconductivity, if the energy of the thermal fluctuations becomes of the order of the Fermi energy then the Fermi surface broadens and the dimensional reduction no more works. 

The thermal fluctuations should destroy the vacuum superconductivity because of the same reason as in the ordinary superconductivity: the loss of the dimensional reduction (Table~\ref{tbl:magnetic}). Indeed, the one--dimensional motion of electric charges in strong magnetic field is 
realized due to the fact that that the electric charges occupy the lowest Landau level which is localized in the transverse plane. The one-dimensional motion can only be spoiled by transitions of the particles to higher, less localized Landau levels. Generally, for a typical gap between the Landau energy levels is expected to be of the QCD scale, $\delta E \sim \Lambda_{\mathrm{QCD}} \approx 100\,\mbox{MeV}$, so that thermal fluctuations of a typical QCD scale, $T \sim \Lambda_{\mathrm{QCD}}$ should destroy the dimensional reduction.

\begin{figure}
\begin{center}
\includegraphics[width=100mm, angle=0]{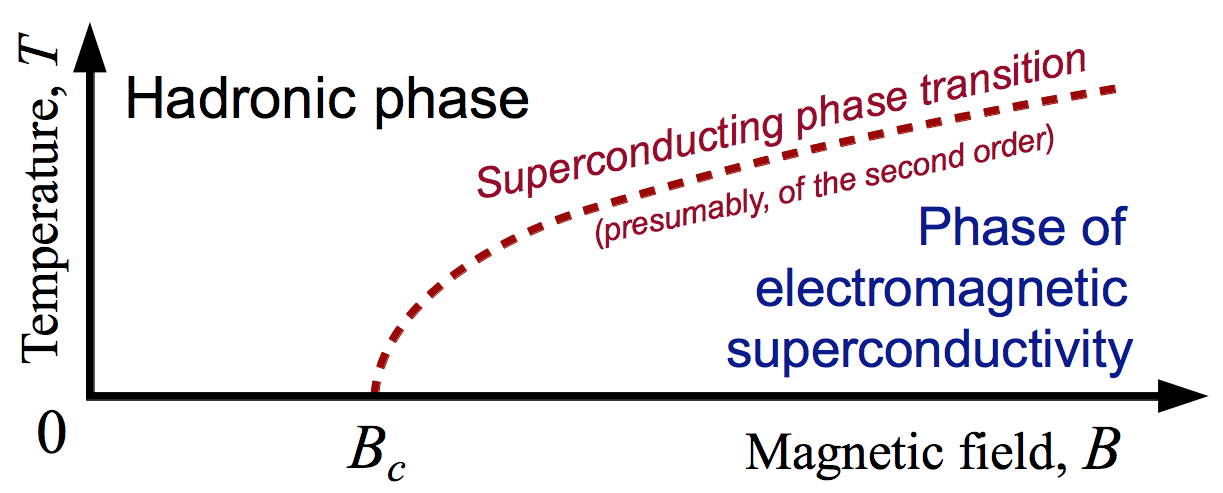} 
\end{center}
\caption{Schematic plot of the suggested QCD phase diagram in the presence of magnetic field in the low temperature region~\cite{Chernodub:2011mc}.}
\label{fig:phase:diagram}
\end{figure}

Therefore, we conclude that the superconductivity should be lost at certain critical temperature, $T_{c} \equiv T_{c}(B)$. At the critical magnetic field, $B=B_{c}$, the critical temperature is zero, $T_{c}(B_{c}) = 0$. The corresponding phase diagram is schematically shown in Fig.~\ref{fig:phase:diagram}: the superconducting and hadronic phases are separated by a phase transition of (presumably) second order~\cite{Chernodub:2011mc}.

\subsubsection{Electric-field-induced pair production (the Schwinger effect) and magnetic--field--induced superconductivity: a comparison}
\label{sec:Schwinger}

The Schwinger effect is a generation of the electron--positron pairs from the vacuum in a background of a strong enough {\emph{electric}} field~\cite{ref:Schwinger}. The created particles form a momentary electric current which tend to screen the external electric field which has created them. The electron--positron pair production is a process which is described entirely by the QED sector of the vacuum. 

The vacuum superconductivity is associated with the emergence of the electrically charged quark-antiquark condensates out of vacuum provided the vacuum is subjected to the strong enough {\emph{magnetic}} field~\cite{Chernodub:2010qx,Chernodub:2011mc}. Contrary to the Schwinger effect, these electrically charged condensates do not screen the external magnetic field which has created them. 

Following Ref.~\cite{Chernodub:2012gr}, in Table~\ref{tbl:Schwinger} we compare the very basic features of the Schwinger effect and the vacuum superconductivity.

\begin{table}[ph]
\caption{Basic features of the Schwinger effect and the electromagnetic vacuum superconductivity.}
\label{tbl:Schwinger}
\begin{tabular}{p{31mm}p{43mm}p{38mm}}
\mline
Property & Schwinger effect  & Vacuum superconductivity \\[1mm]
\mmline
Environment & vacuum &  vacuum \\[1mm]
\nhline
Background of & strong electric field, $E$ & strong magnetic field, $B$ \\[1mm]
\nhline
\multirow{2}{*}{Interactions involved} & electromagnetic (QED) & electromagnetic (QED) \\
                                  &   only                                   & and strong (QCD) \\
\nhline
Typical energy scales & megaelectronvolts ($10^6$ eV) & gigaelectronvolts ($10^9$ eV) \\
\nhline
\multirow{2}{*}{Critical value} & $E_c = m_e^2/e \approx 10^{18}\,\mathrm{V/m}$ & $B_c = m_\rho^2/e \approx 10^{16}\,\mathrm{T} $ \\[1mm]
& {\scriptsize{($m_e = 0.511\,\mbox{MeV}$ is electron mass)}} & {\scriptsize{($m_\rho = 0.775 \,\mbox{GeV}$ is $\rho$--meson mass)}} \\
\nhline
\multirow{3}{*}{Nature of the effect} & virtual electron-positron ($e^- e^+$)  &  virtual quark-antiquark pairs \\
                                                         & pop up from the vacuum  &      ($u \bar{u}$ and $d \bar{d}$) pop up and form \\
  & and become real $e^- e^+$ pairs & real $u \bar{d}$ and $d \bar{u}$ condensates \\
\nhline
\multirow{2}{*}{Backreaction} & created $e^- e^+$ pairs tend  &  created $u \bar{d}$ and $d \bar{u}$ condensates  \\
  & to screen the external field & do not screen the external field\\
\nhline
Stability & a process (unstable)  & a ground state (stable)  \\
\nhline
\multirow{2}{*}{Transport property} & an electromagnetic conductor:  & \multirow{2}{*}{a steady superconducting state}\\
& electric current is generated &    \\
\nhline
\end{tabular}
\end{table}

\subsubsection{Electromagnetic superconductivity of vacuum and ``reentrant superconductivity'' in strong magnetic field}
\label{sec:reentrant}

The magnetic-field-induced electromagnetic superconductivity of vacuum may have counterparts in certain condensed matter systems. It was suggested in Ref.~\cite{ref:Tesanovic} that in a very strong magnetic field the Abrikosov flux lattice of a type-II superconductor may enter a quantum limit of the ``low Landau level dominance'', characterized by a spin-triplet pairing, absence of the Meissner effect, and a superconducting flow along the magnetic field axis. The mentioned quantum limit is reached when  the magnetic length $1/\sqrt{|eB|}$ becomes of the order of the correlation length $\xi$.

In condensed matter, the magnetic-field-induced anisotropic superconductivity is sometimes called the ``reentrant'' superconductivity because the system should normally ``exit'' a superconducting state as an increasing external magnetic field suppresses superconductivity via diamagnetic and Pauli pair breaking effects. Although it is unclear whether this particular mechanism of the reentrant superconductivity works in real superconductors, the magnetic--field--induced restoration of  superconductivity was experimentally observed in certain materials like an uranium superconductor URhGe~\cite{ref:Uranium}.

Our proposal~\cite{Chernodub:2010qx,Chernodub:2011mc} of the vacuum superconductivity has basically the same features as the reentrant superconductivity~\cite{ref:Tesanovic}: the electrically charged condensates correspond to a spin-one quark-antiquark states ($\rho$ mesons), the vacuum superconductor exhibits no Meissner effect while the vacuum superconductivity is highly anisotropic.

\section{Ground state of vacuum superconductor}
\label{sec:quantitative}

\subsection{Energetic favorability of the superconducting state}
\label{sec:energetic}

As we have argued in the previous section, a quark-antiquark pair of different flavors may condense in sufficiently strong magnetic field. How strong should the relevant magnetic field be? Following Ref.~\cite{Chernodub:2010qx}, let us make a simple estimation of the critical magnetic field $B_{c}$ using very general arguments.

An electromagnetic superconductivity emerges when an electrically charged field starts to condense. Assume that we have a free relativistic particle with mass $m$, electric charge $e$ and spin $s$. In a  background of a constant uniform magnetic field $B$ the relativistic energy levels $\varepsilon$ of this particle are given by the following formula:
\beqn
\varepsilon_{n,s_z}^2(p_z) = p_z^2+(2 n - g s_z + 1) |eB| + m^2\,, 
\label{eq:energy:levels}
\eeqn
where $n\geqslant 0$ is the nonnegative integer, $s_z = -s, \dots, s$ is the projection of the spin $s$ on the field's axis, $p_z$ is the particle's momentum along the field's axis and $g$ is the gyromagnetic ratio (or, ``$g$--factor'') of the particle\footnote{Here we ignore the internal structure (formfactors) of the mesons treating them as pointlike bound states. We proceed similarly to the case of the ordinary superconductivity, where the large--sized Cooper pairs can also be treated as local objects in certain approaches, Section~\ref{sec:GL:BCS}, page~\pageref{sec:GL:BCS}. More general and formfactor--independent treatment is presented in Section~\ref{sec:NJL}, page~\pageref{sec:NJL}.}.

Let us consider quark-antiquark bound states made of lightest, $u$ and $d$ quarks. The corresponding simplest spin-zero and spin-one bound states are called $\pi$ and $\rho$ mesons, respectively.

The ground state energy (or, ``mass'') of the $\pi^{\pm}$ mesons in the background of the magnetic field corresponds to the quantum numbers $p_z=0$ and $n_z = 0$ (notice that $s_z \equiv 0$ since the $\pi$ meson is a spineless particle):
\beqn
m_{\pi^\pm}^2(B) = m_{\pi}^2 + |e B|\,,
\label{eq:m2:pi:B:real}
\eeqn
where $m_{\pi} = 139.6$~MeV is the mass of the $\pi$ meson in the absence of the magnetic field. The mass of the neutral $\pi^{0}$ meson is insensitive to the external magnetic field in our approximation, $m_{\pi^{0}}(B) = m_{\pi}$ (here we ignore a small splitting between masses of charged and neutral $\pi$ mesons at $B=0$).

Analogously, the ground state energy (``mass'') of the charged $\rho^{\pm}$ mesons correspond to the quantum numbers $p_z=0$, $n_z = 0$ and $s_z = 1$:
\beqn
m_{\rho^\pm}^2(B) = m_{\rho}^2 - |e B|\,,
\label{eq:m2:pi:B}
\eeqn
where $m_{\rho} = 775.5$~MeV is the mass of the charged $\rho$ meson in the absence of the magnetic field. The mass of the neutral $\rho^{0}$ meson is a $B$--independent quantity in this approximation, $m_{\rho^{0}}(B) = m_{\rho}$ (a small difference in masses of charged and neutral $\rho$ mesons at $B=0$ is ignored again).

\label{page:gyromagnetic}
It is important to mention that in Eq.~\eq{eq:m2:pi:B} the gyromagnetic ratio of the $\rho$ meson is taken to be $g=2$. This anomalously large value was independently obtained in the framework of the QCD sum rules~\cite{ref:g2:sumrules} and in the Dyson--Schwinger approach to QCD~\cite{ref:g2:DS}. It was also conformed by the first-principle numerical simulations of lattice QCD~\cite{ref:g2:lattice} providing us with a value $g \approx 2$. The condensation of the charged $\rho$ mesons in the vacuum of QCD is very similar to the Nielsen-Olesen instability of the pure gluonic vacuum in Yang-Mills theory~\cite{ref:NO} and to the magnetic-field-induced Ambj\o rn--Olesen condensation of the $W$-bosons in the vacuum of standard electroweak model~\cite{ref:AO}. Both the $\rho$ mesons in QCD, the gluons in Yang-Mills theory, and the $W$ bosons in the electroweak model have the anomalously large $g$--factor, $g \approx 2$. Notice, that the phase diagrams of QCD at finite density (at finite chemical and/or isospin potential) contain certain phases characterized by the presence of exotic vector condensates~\cite{ref:Johanna:review,ref:vector:dense1,ref:vector:dense2,ref:vector:dense3}. Some of these phases exhibit superconducting/superfluid behavior~\cite{ref:vector:dense2,ref:vector:dense3}.

In the absence of the magnetic field background the $\rho$ meson is a very unstable particle. One can notice, however, that the ground-state mass~\eq{eq:m2:pi:B:real} of the charged $\pi$ meson is an increasing function of the magnetic field strength $B$ while the mass of the charged $\rho$ mesons~\eq{eq:m2:pi:B} is a decreasing function of $B$. As all known modes of the $\rho^{\pm}$-meson decays proceed via emission of the $\pi^{\pm}$ mesons, $\rho^{\pm} \to \pi^{\pm} X$~\cite{ref:PDG}, it is clear that at certain strength of magnetic field the fast hadronic decays of the $\rho$ mesons become forbidden due to simple kinematical arguments (the mass of the would-be decay products exceed the mass of the $\rho$ meson itself). Thus, in a background of sufficiently strong magnetic field the charged $\rho$ meson should be stable against all known hadronic decay modes~\cite{Chernodub:2010qx}.

The presence of the superconducting ground state at high magnetic field can be seen as follows. The square of mass of the charged $\rho$ meson should decrease as the magnetic field $B$ increases, Eq.~\eq{eq:m2:pi:B}. When the magnetic field reaches the critical value~\eq{eq:Bc}, the ground state energy of the $\rho^\pm$ mesons becomes zero.  As the magnetic field becomes even stronger, the ground state energy becomes a purely imaginary quantity indicating the presence of a tachyonic instability of vacuum. In other words,  the trivial ground state, $\left\langle \rho \right\rangle = 0$, is no more stable at $B > B_{c}$, and the vacuum should slide towards a new state with a nonzero $\rho$--meson condensate, $\left\langle \rho \right\rangle \neq 0$. Since the condensation of the electrically charged field indicates the presence of electromagnetic superconductivity, the vacuum should become a superconductor at $B > B_{c}$.

\subsection{Approaches: Ginzburg-Landau vs Bardeen-Cooper-Schrieffer}
\label{sec:GL:BCS}

The condensation of the $\rho$ mesons in QCD in the background of the strong magnetic field may be treated in the same way as the condensation of the Cooper pairs in the conventional superconductivity. 

The conventional superconductivity may be described in the framework of both microscopic fermionic models and macroscopic bosonic theories. 
The fermionic models of the Bardeen-Cooper-Schrieffer (BCS) type describe basic carriers of electric charge (electrons and/or holes). The BCS models are, generally, nonrenormalizable because their Lagrangians include four-fermion term(s). The bosonic models of the Ginzburg-Landau (GL) type are usually based on renormalizable effective Lagrangians which describe superconducting excitations (the Cooper pairs)~\cite{ref:Abrikosov}. 

Despite of the fact that fermionic and bosonic approaches are formulated in a very different way, they both can describe the superconductivity at a good quantitative level. Moreover, the bosonic and fermionic approaches are mathematically equivalent near the superconducting phase transition~\cite{ref:equivalence}. In Table~\ref{tbl:models} we outline a correspondence between the traditional (BCS and GL) models of conventional superconductivity and their vacuum counterparts which will be used to describe the magnetic--field--induced vacuum superconductor later. 

\begin{table}[ph]
\caption{Simplest models which are used to study physical properties of conventional and vacuum superconductivities.}
\label{tbl:models}
\begin{tabular}{p{36mm}p{30mm}p{46mm}}
\mline
Basic field describes ... & conventional  & vacuum  \\[1mm]
\nhline
bosonic, condensed & Ginzburg--Landau & $\rho$-meson electrodynamics~\cite{Djukanovic:2005ag} based  \\[1mm]
superconducting carriers &  model~\cite{ref:GL} & on vector dominance model~\cite{Sakurai:1960ju} \\[1mm]
\nhline
fermionic constituents of &  Bardeen--Cooper-- & Nambu--Jona-Lasinio model~\cite{ref:NJL} \\[1mm] 
superconducting carriers & Schrieffer model~\cite{ref:BCS} & extended with vector interactions~\cite{Ebert:1985kz} \\[1mm]
\end{tabular}
\end{table}

It is important to notice that in the effective GL approach the Cooper pairs are treated as pointlike particles. However, physically the Cooper pairs are rather nonlocal objects because their size is much larger than the average distance between electrons in metals. Nevertheless, the GL model describes superconductivity very well especially near the second--order phase transition where the symmetries of the system dominate its dynamics according to the universality argument. 

Following our experience in the conventional superconductivity, we describe below a GL like approach to the vacuum superconductivity using a model of a  $\rho$-meson sector of vacuum~\cite{Djukanovic:2005ag}. Then, we briefly outline a Bardeen-Cooper-Schrieffer approach to the vacuum superconductor using a well-known extension~\cite{Ebert:1985kz} of the Nambu--Jona-Lasinio model~\cite{ref:NJL}. We would like also to mention that signatures of the vacuum superconductor were also found in holographic effective theories~\cite{Callebaut:2011ab} and in numerical (``lattice'') approaches to QCD~\cite{Braguta:2011hq}.

\subsection{Example: Ginzburg--Landau model}
\label{sec:GL}

Before going into the details of the $\rho$ condensation in QCD, it is very useful to outline a few basic properties of the conventional superconductivity in the GL model.

\subsubsection{The relativistic version of the Ginzburg--Landau Lagrangian}

Conventional superconductors can be described by the following GL Lagrangian:
\beqn
\cL_{\mathrm{GL}} = - \frac{1}{4} F_{\mu\nu} F^{\mu\nu} + (\cD_\mu \Phi)^* \cD^\mu \Phi - \lambda (|\Phi|^2 - \eta^2)^2\,, \quad
\label{eq:L:GL}
\eeqn
where $\cD_\mu = \partial_\mu - i e A_\mu$ is the covariant derivative and $\Phi$ is the complex scalar field carrying the electric charge\footnote{Without loss of generality we consider the singly-charged bosons $\Phi$ instead of the doubly charged Cooper pairs and we use a relativistic description of superconductivity.}~$e$.

The superconducting ground state of the GL model, $\eta^{2} > 0$, is characterized by the presence of the homogeneous condensate $\langle \Phi\rangle$, with $|\langle \Phi\rangle| = \eta$, which is responsible for the superconductivity.  The condensate breaks the electromagnetic symmetry group of Lagrangian~\eq{eq:L:GL}, $\Phi \to e^{i e \omega} \Phi$ and $A_{\mu} \to A_{\mu} + \partial_{\mu} \omega$.

In the superconducting phase the mass of the scalar excitation, $\delta \Phi = \Phi - \langle \Phi\rangle$, and the mass of the photon field $A_\mu$ are, respectively, as follows:
\beqn
m_\Phi = \sqrt{4 \lambda} \eta\,,
\qquad \
m_A = \sqrt{2} e \eta\,.
\label{eq:masses}
\eeqn
The classical equations of motion of the GL Lagrangian~\eq{eq:L:GL} are:
\beqn
\cD_\mu \cD^\mu \Phi + 2 \lambda (|\Phi|^2 - \eta^2) \Phi & = & 0\,,
\label{eq:GL:1}\\
\partial_\nu F^{\nu\mu} + J_{\mathrm{GL}}^\mu & = & 0\,,
\label{eq:GL:2}
\eeqn
where the electric current is
\beqn
J_{\mathrm{GL}}^\mu = - i e \bigl[\Phi^* \cD^\mu \Phi - (\cD^\mu \Phi)^* \Phi \bigr]\,.
\label{eq:GL:J}
\eeqn

Thermal fluctuations make the condensate $\langle \Phi\rangle$ smaller, eventually destroying the superconductivity at certain critical temperature, $T = T_{c}$. In order to describe this effect in the GL approach, one usually assumes that the quadratic coefficient of the potential term $V_{\mathrm{GL}} = \lambda (|\Phi|^2 - \eta^2)^2$ in the GL Lagrangian~\eq{eq:L:GL} exhibits a linear temperature dependence:
\beqn
V_{\mathrm{GL}} (\Phi) = \alpha_{0} (T - T_{c}) |\Phi|^2 + \lambda |\Phi|^4 + {\mathrm{const.}} \quad
\label{eq:V:GL}
\eeqn
The superconducting condensate is present at $T<T_{c}$, while $\langle \Phi\rangle = 0$ at $T \geqslant T_{c}$.

\subsubsection{Magnetic field destroys conventional superconductivity}
\label{sec:GL:destructive}

In a background of a sufficiently strong magnetic field, $B > B^{\mathrm{GL}}_c$, the superconducting condensate disappears. The corresponding critical value of the magnetic field,
\beqn
B^{\mathrm{GL}}_c = \frac{m^2_\Phi}{2e} \equiv \frac{2 \lambda}{e} \eta^2\,,
\label{eq:B:GL:c}
\eeqn
can be obtained as follows. Consider a near-critical case when the uniform time-independent magnetic field $B \equiv F_{12}$ is slightly weaker than the critical value~\eq{eq:B:GL:c},  $0 < B^{\mathrm{GL}}_c - B \ll B^{\mathrm{GL}}_c$. As the magnetic field approaches the critical value~\eq{eq:B:GL:c}, the superconducting condensate becomes very small,
\beqn
|\langle \Phi\rangle(B)| \ll \eta\,,
\label{eq:first}
\eeqn
and, consequently,  the classical equation of motion~\eq{eq:GL:1} can be linearized:
\beqn
\bigr\{(\cD_1 - i \cD_2) (\cD_1 + i \cD_2) + e [B_c^{\mathrm{GL}} - B(x)]\bigl\} \Phi = 0\,.
\label{eq:D1D2}
\eeqn
The magnetic field inside the superconductor, $B(x)$, appears explicitly in Eq.~\eq{eq:D1D2} due to a rearrangement of the derivatives of the first term.

In the vicinity of the critical magnetic field, $B \simeq B_c$, so that the second term in Eq.~\eq{eq:D1D2} can be neglected and Eq.~\eq{eq:D1D2} is satisfied if 
\beqn
\cD \Phi \simeq 0 \qquad \mbox{with} \quad \cD = \cD_1 + i \cD_2\,.
\label{eq:cD:phi}
\eeqn

\subsubsection{Lattice of Abrikosov vortices in background of magnetic field}
\label{sec:GL:Abrikosov}

If the strength of the external magnetic field is smaller then the critical value~\eq{eq:B:GL:c}  then the superconductor may squeeze the magnetic field into the vortexlike structures which are known as the Abrikosov vortices~\cite{Abrikosov:1956sx}. The Abrikosov vortex is a topological stringlike solution to the classical equations of motion~\eq{eq:GL:1} and \eq{eq:GL:2}. A single Abrikosov vortex carries a quantized flux of the magnetic field,
\beqn
\int \dd^2 x^{\perp} \, B(x^{\perp}) = \frac{2 \pi}{e}\,,
\label{eq:quantized}
\eeqn
where the integral of the vortex magnetic field $B$ is taken over the two-dimensional coordinates $x^{\perp} = (x^{1},x^{2})$ of the plane which is transverse to the infinitely-long, strait and static vortex [notice, that the flux~\eq{eq:quantized} is twice larger than a conventional value since we consider the condensed bosons $\Phi$ with the electric charge $e$ and not $2e$]. 

The Abrikosov vortex has a well-defined center where the condensate $\Phi$ vanishes restoring the normal (nonsuperconducting) phase. At the vortex center the phase of the scalar field is singular. The behavior of the scalar field in the vicinity of the elementary vortex, situating at the origin and carrying the flux~\eq{eq:quantized}, is as follows 
\beqn
\Phi (x^\perp) \propto |x^{\perp}| e^{i \varphi} \equiv x_1 + i x_2\,,
\label{eq:Phi:vort}
\eeqn
where $\varphi$ is the azimuthal angle in the transverse plane, and $|x^{\perp}|$ is the distance from the vortex center.  Equation~\eq{eq:Phi:vort} is valid provided  $m_\Phi |x^\perp| \ll 1$ and $m_A|x^\perp| \ll 1$, where the mass parameters $m_{A}$ and $m_{\Phi}$ are given in Eq.~\eq{eq:masses}.

If the external magnetic field is strong enough but still weaker than the critical value \eq{eq:B:GL:c}, then multiple elementary Abrikosov vortices may be created. Parallel Abrikosov vortices repel each other in a type--II superconductor, for which the mass of the scalar excitation is larger then the photon mass, $m_\Phi > m_A$.  Due to the mutual repulsion, these vortices arrange themselves in a regular periodic structure known as the Abrikosov lattice~\cite{ref:Abrikosov}. The Abrikosov lattice corresponds to the so called ``mixed state'' of the conventional superconductor, in which both normal phase (inside the vortex cores) and superconducting phase (outside the vortex cores) coexist.

There are various arrangements of the Abrikosov vortices corresponding to different types of Abrikosov lattices~\cite{ref:Abrikosov}. The simplest Abrikosov lattice is given by the following solution of Eq.~\eq{eq:cD:phi}:
\beqn
\Phi & = &\Phi_0 \, K\bigl(z/L_B\bigr)\,, \quad \qquad
K(z) = e^{-\frac{\pi}{2} (|z|^2 + z^2)} \sum\limits_{n = -\infty}^{+ \infty} e^{- \pi n^2 + 2 \pi n z}\,,
\label{eq:K}
\label{eq:Abrikosov:lattice}
\label{eq:v}
\eeqn
where $\Phi_0$ is a dimensional complex parameter, $z = x^1 + i x^2$, and
\beqn
L_B = \sqrt{2 \pi} \cl_B\,, \qquad \ \cl_B = \frac{1}{\sqrt{e B}}\,,
\label{eq:LB:GL}
\eeqn
is the inter-vortex distance $L_B$ expressed via the magnetic length $\cl_B$. The area of an elementary lattice cell (i.e. of a cell which contains one Abrikosov vortex) is $L^2_B$. In the solution~\eq{eq:Abrikosov:lattice}  the vortices are located at the sites of the square lattice,
\beqn
\frac{x_i}{L_B} = n_i + \frac{1}{2}\,, \qquad n_i \in {\mathbb{Z}}\,, \quad i=1,2\,,
\label{eq:vort:locations}
\eeqn
at which the condensate $\Phi(x_1,x_2)$ vanishes exactly. In the vicinity of these sites the scalar field~\eq{eq:Abrikosov:lattice} is described by Eq.~\eq{eq:Phi:vort}.

The ground state of the system in the mixed phase is characterized, by definition, by a minimal energy of the vortex lattice. One can show that a global minimum of the energy density of the GL model~\eq{eq:L:GL} corresponds to a global minimum of a convenient dimensionless quantity which is called the Abrikosov ratio~\cite{ref:Abrikosov}:
\beqn
\beta_{A} = \frac{\langle |\phi|^4 \rangle}{\langle |\phi|^2 \rangle^2}\,.
\label{eq:beta}
\eeqn  

\begin{figure}
\begin{center}
\begin{tabular}{ll}
\includegraphics[width=45mm, angle=0]{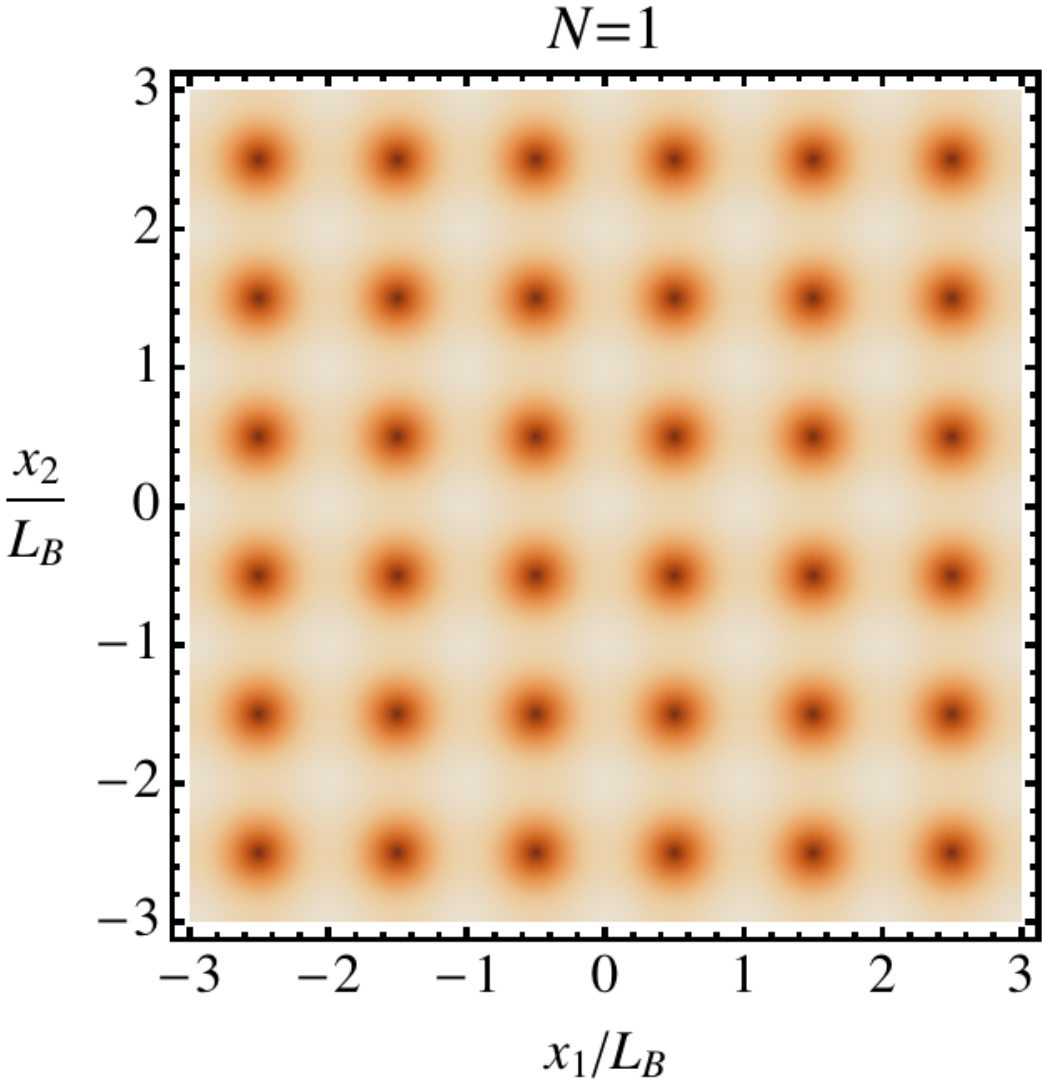} 
& \hskip 5mm
\includegraphics[width=45mm, angle=0]{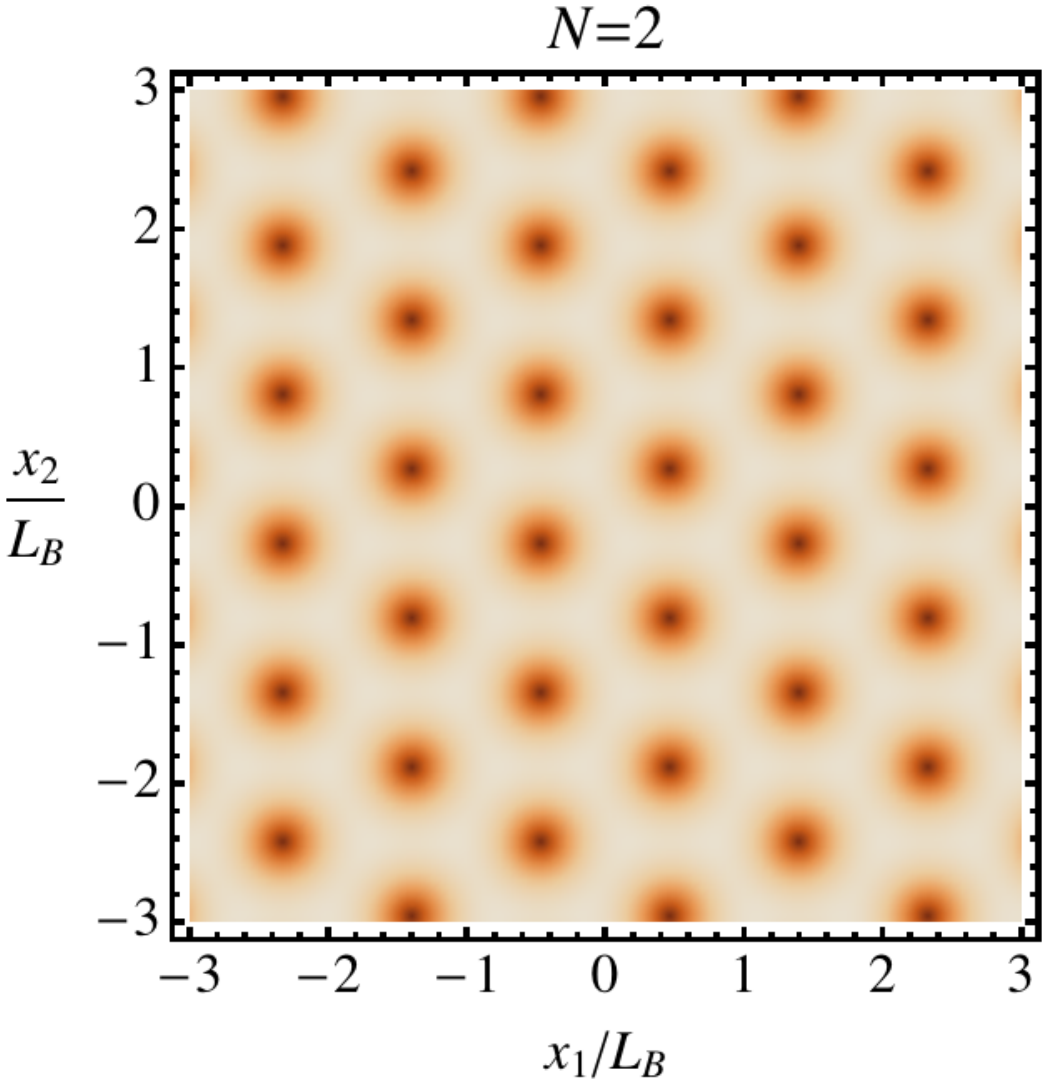}
\end{tabular}
\end{center}
\caption{Minimal-energy vortex lattices for square (left) and equilateral triangular (right) lattices in the Ginzburg--Landau model. The dark dots correspond to the positions of the Abrikosov vortices. The equilateral triangular lattice corresponds to the global minimum of the energy. From Ref.~\cite{ref:Jos}.}
\label{fig:GL:lattices}
\end{figure}

It turns out that the real ground state of the system corresponds an equilateral triangular lattice (which is sometimes also called ``hexagonal'' lattice) with the Abrikosov ratio $\beta_{A}({\mathrm{Triangular}}) \approx 1.1596$, Fig.~\ref{fig:GL:lattices}(right). For the square Abrikosov lattice the Abrikosov ratio~\eq{eq:beta} is slightly higher, $\beta_{A}({\mathrm{Square}}) \approx 1.180$, Fig.~\ref{fig:GL:lattices}(left). Notice that despite very different visual appearances of these two lattices, the difference in their energies (and in the corresponding Abrikosov ratios, $\beta_{A}$) is of the order of a few percent. A Ginzburg--Landau description of the type--II superconductors in the background of magnetic field can be found in the nice review~\cite{ref:type-II:Review}.

\subsubsection{London equations and complex electric conductivity}
\label{sec:GL:homogeneous}

The GL model~\eq{eq:L:GL} in the condensed phase describes a superconductor. Indeed, taking into account that in the ground state the condensate is a uniform time-independent quantity, one finds from from the definition of the electric current~\eq{eq:GL:J}: 
\beqn
\partial^{\mu} J_{\mathrm{GL}}^{\nu} - \partial^{\nu} J_{\mathrm{GL}}^{\mu} = - m^2_A F^{\mu\nu}\,,
\label{eq:dJ:London}
\eeqn
where $m_A$ is given in Eq.~\eq{eq:masses}. Setting $\mu = 0$ and $\nu = 1,2,3$, in Eq.~\eq{eq:dJ:London} we recover the first London relation for a locally neutral [$J_0(x) = 0$] superconductor:
\beqn
\frac{\partial {\vec J}_{\mathrm{GL}}}{\partial t} = m_A^2 {\vec E}\,,
\label{eq:London:GL}
\eeqn
where ${\vec{E}}$ or $E^i \equiv - F^{0i}$ is a weak external electric field. Equation~\eq{eq:London:GL} describes a linear growth of the electric current in the constant electric field, thus indicating a vanishing electric resistance of the system. According to Eq.~\eq{eq:London:GL}, the superconductivity is homogeneous (coordinates-independent) and isotropic (direction-independent).

The London equation~\eq{eq:London:GL} corresponds to a singular part of the complex conductivity tensor $\sigma_{kl} = {\mathrm{Re}}\, \sigma_{kl} + i \, {\mathrm{Im}}\, \sigma_{kl}$. The conductivity tensor is defined as follows:
\beqn
J_k(\vec x, t;\omega) = \sum_{k=1}^3 \sigma_{kl}(\omega) E_l(\vec x, t)\,,
\label{eq:J:London}
\eeqn
where ${\vec E}(\vec x, t) = {\vec E}_0 e^{i (\vec x \cdot \vec q - \omega t)}$ is the alternating external current in the long-wavelength limit, $|\vec q | \to 0$.
The London equation~\eq{eq:London:GL} indicates that $\sigma_{kl}(\omega) = \sigma^{\mathrm{sing}}_{kl}(\omega) + \sigma^{\mathrm{reg}}_{kl}(\omega)$,
where the singular part comes from the paired (superconducting) electrons,
\beqn
\sigma^{\mathrm{sing}}_{kl} (\omega) = \frac{\pi m^2_A}{2} \Bigl[\delta(\omega) + \frac{2 i}{\pi \omega}\Bigr] \delta_{kl}\,,
\label{eq:sigma:London:2}
\eeqn
while the regular part $\sigma^{\mathrm{reg}}_{kl}$ accounts for other contributions to the conductivity.

\subsubsection{Meissner effect}
\label{sec:GL:Meissner}

The spatial components of Eq.~\eq{eq:dJ:London} provide us with the second London relation:
\beqn
{\vec \partial} \times {\vec J}_{\mathrm{GL}} = - m^2_A {\vec B}\,.
\label{eq:London:GL:2}
\eeqn
In the absence of a background electric field ($\vec E = 0$),  Eq.~\eq{eq:GL:2} implies ${\vec J}_{\mathrm{GL}} = {\vec \partial} \times {\vec B}$, so that Eq.~\eq{eq:London:GL:2} can now be reformulated as follows:
\beqn
(- \Delta + m^2_A) {\vec B} = 0\,.
\label{eq:Meissner}
\eeqn

Equation \eq{eq:Meissner} describes the Meissner effect: inside a superconductor the photon becomes a massive particle so that an external magnetic field $B < B_c$ is expelled.  Physically, the Meissner effect appears because the external magnetic field induces circulating superconducting currents~\eq{eq:London:GL:2} which, in turn, generate their own magnetic field. As the generated field is directed in the opposite direction with respect to the external field, the magnetic field is eventually screened inside superconductor. 

If the external magnetic field is directed tangentially with respect the superconductor's boundary then this field is always screened inside the bulk of the superconductor. However, if the external magnetic field is directed along a normal of the boundary of a type--II superconductor, then the magnetic flux -- which is a conserved quantity -- may penetrate the superconductor and create a mixed phase of the Abrikosov vortices (Section~\ref{sec:GL:Abrikosov}, page~\pageref{sec:GL:Abrikosov}).

\subsection{Superconductivity of vacuum in strong magnetic field}
\label{sec:VDM}

\subsubsection{Electrodynamics of $\rho$ mesons }

The conventional superconductivity is driven by the condensation of the Cooper pairs which are described by the local scalar field $\Phi$ in the Ginzburg--Landau approach.  The superconductivity of vacuum in a sufficiently strong magnetic field is caused by emergence of quark-antiquark condensates which carry quantum numbers of (charged) $\rho$ mesons~\cite{Chernodub:2010qx}. In order to describe the electrodynamics of the $\rho$ mesons in the GL style, we use the following effective Lagrangian~\cite{Djukanovic:2005ag}:
\beqn
{\cal L} & = & -\frac{1}{4} \ F_{\mu\nu}F^{\mu\nu}
- \frac{1}{2} (D_{[\mu,} \rho_{\nu]})^\dagger D^{[\mu,} \rho^{\nu]} + m_\rho^2 \ \rho_\mu^\dagger \rho^{\mu} \nonumber
\\ & & 
 - \frac{1}{4} \rho^{(0)}_{\mu\nu} \rho^{(0) \mu\nu}{+}\frac{m_\rho^2}{2} \rho_\mu^{(0)}
\rho^{(0) \mu} +\frac{e}{2 g_s} F^{\mu\nu} \rho^{(0)}_{\mu\nu}\,, 
\label{eq:L} 
\eeqn
where the complex vector field $\rho_\mu = (\rho^{(1)}_\mu - i \rho^{(2)}_\mu)/\sqrt{2}$ and the real-valued vector field $\rho^{(0)}_\mu \equiv \rho^{(3)}_\mu$, correspond, respectively, to the charged and neutral vector mesons made of the components of the triplet of the $\rho$ field:
\beqn
\rho_{\mu} = 
\left(
\rho_{\mu}^{(1)},
\rho_{\mu}^{(2)},
\rho_{\mu}^{(3)} 
\right)^{T}\,.
\label{eq:rho:field}
\eeqn

The last term in Eq.~\eq{eq:L} describes a nonminimal coupling of the $\rho$ mesons to the electromagnetism via the field strength $F_{\mu\nu} = \partial_{[\mu,} A_{\nu]}$ of the photon field $A_\mu$. The presence of the nonminimal coupling implies, in particular, the anomalously large value of the gyromagnetic ratio of the $\rho$ meson, $g = 2$ (discussed already in Section~\ref{sec:energetic}, page \pageref{page:gyromagnetic}). Both the covariant derivative $D_\mu = \partial_\mu + i g_s \rho^{(0)}_\mu - ie A_\mu$ and the strength tensor $\rho^{(0)}_{\mu\nu} = \partial_{[\mu,} \rho^{(0)}_{\nu]} - i g_s \rho^\dagger_{[\mu,} \rho_{\nu]}$ involve the $\rho\pi\pi$ coupling $g_s$ which has the known phenomenological value of $g_s \approx 5.88$, Ref.~\cite{Djukanovic:2005ag}. 

Lagrangian~\eq{eq:L} is invariant under the electromagnetic gauge transformations, 
\beqn
\rho_\mu(x) \ \  \to e^{i e \omega(x)} \rho_\mu(x)\,, \qquad
A_\mu(x) \ \, & \to & A_\mu(x) + \partial_\mu \omega(x)\,,
\label{eq:gauge:transformations}
\eeqn
which do not affect the neutral field, $\rho^{(0)}_\mu(x) \to \rho^{(0)}_\mu(x)$.

The $\rho$--meson  Lagrangian~\eq{eq:L}  is an analogue of the Ginzburg--Landau Lagrangian~\eq{eq:L:GL}, while the $\rho$--meson field~\eq{eq:rho:field} plays the role of the GL scalar field $\Phi$. The electric current of the $\rho$ mesons is given by the analogue of Eq.~\eq{eq:GL:J}:
\beqn
J_\mu = i e \bigl[\rho^{\nu\dagger} \rho_{\nu\mu} - \rho^\nu \rho^\dagger_{\nu\mu} 
+ \partial^\nu (\rho^\dagger_\nu \rho_\mu - \rho^\dagger_\mu \rho_\nu)\bigr] - \frac{e}{g_s} \partial^\nu f^{(0)}_{\nu\mu}\,.
\label{eq:Jmu:2}
\eeqn

\subsubsection{Instability of vacuum: potential energy in strong magnetic field}

The energy density ${\cal E}$ of the $\rho$--meson ground state is given by the $T_{00}$ component,
\beqn
& & \hskip -4mm {\cal E} \equiv T_{00} = \frac{1}{2} F_{0i}^2 + \frac{1}{4} F_{ij}^2
+ \frac{1}{2} \bigl(\rho_{0i}^{(0)}\bigr)^2 + \frac{1}{4} \bigl(\rho_{ij}^{(0)}\bigr)^2
+ \frac{m_\rho^2}{2}\Bigl[\bigl(\rho_{0}^{(0)}\bigr)^2 + \bigl(\rho_{i}^{(0)}\bigr)^2\Bigr]
\nonumber\\
& & 
+ \rho_{0i}^\dagger \rho_{0i} + \frac{1}{2} \rho_{ij}^\dagger \rho_{ij}
+ m_\rho^2\bigl(\rho_{0}^\dagger \rho_{0} + \rho_{i}^\dagger \rho_{i}\bigr)
- \frac{e}{g_s} F_{0i} \rho^{(0)}_{0i} - \frac{e}{2 g_s} F_{ij} \rho^{(0)}_{ij}\,,
\label{eq:energy:density}
\eeqn
of the energy-momentum tensor of the $\rho$--meson electrodynamics~\eq{eq:L}:
\beqn
T_{\mu\nu} = 2 \frac{\partial {\cal L}}{\partial g^{\mu\nu}} - {\cal L}\, g_{\mu\nu}\,.
\eeqn

It is useful to consider a ``homogeneous'' approximation and ignore for a moment all derivatives and covariant derivatives in the energy density~\eq{eq:energy:density}. This procedure corresponds, roughly speaking, to selection of a potential part of the energy density:
\beqn
V \left( \rho_{\mu}, \rho_{\mu}^{(0)}\right) & = & \frac{1}{2} B^2 + \frac{g_s^2}{4} \sum_{\mu,\nu=0}^{4}
\bigl[i \bigl(\rho_\mu^\dagger \rho_\nu - \rho_\nu^\dagger \rho_\mu\bigr)\bigr]^2
+ i e B \bigl(\rho_1^\dagger \rho_2 - \rho_2^\dagger \rho_1\bigr)\label{eq:epsilon0} 
\nonumber \\ & &  
+ \frac{m_\rho^2}{2} \sum_{\mu = 0}^{4} {\bigl(\rho_\mu^{(0)}\bigr)}^2  
+ m_\rho^2  \sum_{\mu = 0}^{4} \rho_\mu^\dagger \rho_\mu\,. \quad
\label{eq:V:homogeneous}
\eeqn
In the homogeneous approximation the ground state can be found via the minimization of the potential energy~\eq{eq:epsilon0} with respect to the fields $\rho_{\mu}$ and $\rho_{\mu}^{(0)}$. It turns out that in this approximation the vacuum expectation value of the neutral $\rho$-meson field is zero, $\rho_\mu^{(0)}=0$. The quadratic part of the charged field is the following:
\beqn
V^{(2)}(\rho_\mu) = \sum_{a,b=1}^2\rho_a^\dagger \cM_{ab} \rho_b + m_\rho^2 (\rho_0^\dagger \rho_0 + \rho_3^\dagger \rho_3)\,,
\label{eq:rho:mu}
\qquad
\cM =
\left(
\begin{array}{cc}
m_\rho^2 & i e B \\
- i e B & m_\rho^2
\end{array}
\right)\,, \quad
\label{eq:cM}
\eeqn
where the mass matrix $\cM$ for the Lorentz components $\rho_1$ and $\rho_2$ is non--diagonal.

The eigenvalues $\mu_{\pm}$ and the corresponding eigenvectors $\rho_{\pm}$
of the mass matrix \eq{eq:cM} are, respectively, as follows:
\beqn
\mu_{\pm}^2 = m_\rho^2 \pm e B\,,
\qquad
\rho_{\pm} = \frac{1}{\sqrt{2}} (\rho_1 \mp i \rho_2)\,.
\label{eq:rho:diag}
\eeqn
It is clearly seen that one of mass states, either $\mu_{-}$ or $\mu_{+}$ depending on the sign of $e B$, is getting smaller as the magnetic field increases. Taking for definiteness $e B > 0$, we chose the ground state of the system in the following form:
\beqn
\rho_1 = \rho\,, \qquad \rho_2 = - i \rho\,, \qquad \rho_0 = 0\,, \qquad \rho_3 = 0\,.
\label{eq:ground:state}
\eeqn
The longitudinal components $\rho_0$ and $\rho_3$ are always zero because for any value of the magnetic field the corresponding terms in Eq.~\eq{eq:rho:mu} are positive and diagonal.

The total potential energy of the $\rho$ meson system in the homogeneous ground state can be calculated with the help of Eqs.~\eq{eq:V:homogeneous} and \eq{eq:ground:state}:
\beqn
V(\rho) = \frac{1}{2} B^2 + 2 e (B_{c} - B) \, |\rho|^2 + 2 g_s^2 \, |\rho|^4\,, 
\label{eq:epsilon:0:2}
\eeqn
where 
\beqn
B_{c} = \frac{m_\rho^2}{e}\,,
\eeqn
is the critical magnetic field~\eq{eq:Bc}.

Thus, we get the familiar Mexican-hat potential~\eq{eq:epsilon:0:2} for the $\rho$--meson condensate $\rho$. In particular, the very same form of the potential appears in the GL model of superconductivity~\eq{eq:V:GL}, with one very important exception: in the conventional superconductivity the condensation of the Cooper pairs emerges at low temperatures, $T < T_{c}$, while the $\rho$ mesons start to condense in the vacuum in the presence of sufficiently strong magnetic fields, $B > B_{c}$:
\beqn
|\langle \rho \rangle |_{V} =
\left\{
\begin{array}{lcl}
\sqrt{\frac{e(B - B_c)}{2 g^2_s}}\,, & \quad & B  \geqslant B_c\,,\\
0\,, & \quad & B < B_c\,.
\end{array}
\right.
\label{eq:norm:rho2:h}
\eeqn
Here the subscript $V$ indicates that we consider the potential part $V(\Phi)$ only so that we ignore all kinetic terms. If $B  \geqslant B_c$ then the condensate~\eq{eq:norm:rho2:h}  breaks spontaneously the electromagnetic symmetry group~\eq{eq:gauge:transformations}, similarly to the spontaneous symmetry breaking in the superconducting phase of the GL model.

\subsubsection{Negative condensation energy due to $\rho$--meson condensate}

The homogeneous nature of the $\rho$--meson condensate~\eq{eq:norm:rho2:h} is an artifact of the potential approximation which was useful to find the very presence of the tachyonic instability of the noncondensed state at $B > B_{c}$. In order to determine a detailed structure of the ground state, we should work beyond the potential approximation. To this end, we notice that a wavefunction of the lowest energy state of a free particle in a uniform static magnetic field is independent on the coordinate $z \equiv x^{3}$ along the magnetic field axis. Secondly, the dependence on the time coordinate $t \equiv x^{0}$ should appear only in a form of a trivial phase factor. Thus, we concentrate on $x^{1}$- and $x^{2}$-dependent solutions to the classical equation of motions for the $\rho$ mesons, similarly to the case of the ordinary Abrikosov lattice solutions. 

Technically, it is convenient to choose the complex coordinate $z = x^{1} + i x^{2}$ and define the complex variables ${\cal O} = {\cal O}_1 + i {\cal O}_2$ and their conjugates ${\overline{\cal O}} = {\cal O}_1 - i {\cal O}_2$ for vector quantities ${\cal O}_{\mu}$. Then, the classical equations of motion of the $\rho$--meson model~\eq{eq:L} can be written in the following ``complexified'' form:
\beqn
g_s \partial B + i e m^2_0 \rho^{(0)} & = & 0,
\label{eq:one} \\
\bigl( - \bar{\partial}\partial + m^2_0 + 2 g_s^2 |\rho|^2\bigr) \rho^{(0)} - 2 i g_s \partial |\rho|^2 & = & 0, \qquad
\label{eq:two}\\
\bigl[- \bar{D} D + 2 \bigl(g_s C - e B + 2 g_s^2 |\rho|^2 + m_\rho^2\bigr)\bigr] \rho & = & 0\,,
\label{eq:three}
\eeqn
where the $z$--components of the magnetic field and its analogue for the neutral $\rho$ mesons are, respectively, as follows
\beqn
B(z)  \equiv  \partial_{1} A_{2} - \partial_{2} A_{1} = {\mathrm{Im}}(\bar{\partial} A)\,,
\label{eq:B:2d}
\qquad
C(z) \equiv \partial_{1} \rho_{2}^{(0)} - \partial_{2} \rho_{1}^{(0)} = {\mathrm{Im}}(\bar{\partial} \rho^{(0)})\,. 
\label{eq:C:2d}
\eeqn
We have also introduced two covariant derivatives:
\beqn
D \equiv D_1 + i D_2 = \cD + i g_s \rho^{(0)}\,, \qquad \cD = \partial - i e A\,, \qquad A = \frac{B_{\ext}}{2 i} z\,,
\label{eq:covariant}
\eeqn
where the external uniform magnetic field $B_{\ext}$ should be distinguished from the full magnetic field in the superconducting state, $B(x_{1},x_{2}) \equiv B(x_1 + i x_2)$ as the latter induces a backreaction from the superconducting ground state.

The transverse components of the electric current~\eq{eq:Jmu:2} are as follows:
\beqn
J^{\perp} \equiv J_{1} + i J_{2} = 2 i e \bigl(\rho^\dagger D \rho + \partial |\rho|^2 \bigr) + i \frac{e}{g_s}\partial C\,.
\label{eq:J:perp}
\eeqn

In the vicinity of the phase transition, $B_{\ext} > B_{c}$ with $|B_{\ext} - B_{c}| \ll B_{c}$, the equations of motion \eq{eq:one}, \eq{eq:two} and \eq{eq:three} can be linearized. It turns out that the equation for the $\rho$-meson condensate in the overcritical magnetic field ($B_{\ext} \gtrsim B_c$) in the vacuum is identical to the equation for the Cooper pair condensate~\eq{eq:cD:phi} in the subcritical magnetic field ($B_{\ext} \lesssim B_c$) in the GL model of conventional superconductivity~\cite{Chernodub:2010qx}:
\beqn
{\mathfrak D} \rho \equiv (\partial -\frac{e}{2} B_\ext z) \rho = 0\,.
\label{eq:cDrho0}
\eeqn
Among infinite number of solutions to Eq.~\eq{eq:cDrho0}, the ground state solution corresponds to the global energy minimum. In the chosen approximation, the mean value of the energy density~\eq{eq:energy:density} can be expressed via the $\rho$--meson condensate~\cite{ref:Jos}:
\beqn
\langle {\cal E} \rangle \equiv \langle T_{00} \rangle & = & \frac{1}{2} B_\ext^2 + 2 e (B_{c} - B_\ext) \langle |\rho|^2\rangle 
+ 2 e^2 \langle |\rho|^2\rangle^2  \nonumber \\
& & + 2 \bigl(g_s^2 - e^2\bigr) \left\langle \vert \rho\vert^2\frac{m^2_0}{ - \Delta + m^2_0}\vert \rho\vert^2\right\rangle \,,
\label{eq:E:rho}
\eeqn
where $\partial^2_{\perp} \equiv \partial^2_1 + \partial^2_2$ is the two--dimensional Laplace operator, 
\beqn
\frac{1}{- \partial^2_{\perp} + m^2_0}(x^\perp) = \frac{1}{2 \pi} K_0(m_{0}|x^\perp|)
\eeqn
is a two-dimensional Euclidean propagator of a scalar massive particle with the mass of the neutral $\rho^{(0)}$ meson,
\beqn
m_0 \equiv m_{\rho^{(0)}} = m_\rho \Bigl(1 - \frac{e^2}{g_s^2}\Bigr)^{-\frac{1}{2}}\,,
\label{eq:m:rho0}
\eeqn
and $K_0$ is a modified Bessel function. Contrary to the potential part of the energy density~\eq{eq:epsilon:0:2}, the full expression \eq{eq:E:rho} depends nonlocally on the condensate $\rho$.

A general solution of Eq.~\eq{eq:cDrho0} is the following generalization to the square Abrikosov lattice~\eq{eq:Abrikosov:lattice}:
\beqn
\rho(z) = \sum_{n \in \Z} C_n \exp\Bigl\{ - \frac{\pi}{2} \bigl(|z|^2 + {\bar z}^2\bigr) - \pi \nu^2 n^2 + 2 \pi \nu n {\bar z}\Bigr\}\,, \qquad z = x_{1} + i x_{2}\,,
\label{eq:rho:z}
\eeqn
where $\nu$ is an arbitrary real parameter, $L_B$ is the magnetic length~\eq{eq:LB:GL} and $C_{n}$ are arbitrary complex coefficients.

The ground state solution is given by an equilateral triangular lattice, Fig.~\ref{fig:GL:lattices}(right). The corresponding coefficients $C_{n}$ obey the two--fold symmetry $C_{n+2} = C_{n}$ with $C_{1} = i C_{0}$, while the independent parameters $\nu$ and $C_{0}$ cannot be calculated analyticity so that they should be found by a numerical minimization of the energy density~\eq{eq:E:rho}. Equivalently, one can also minimize an analogue the Abrikosov ratio~\eq{eq:beta}:
\beqn
\beta_{\rho} = \left\langle \frac{\vert \rho \vert^2}{\langle |\rho|^2 \rangle} \frac{m^2_0}{ - \Delta + m^2_0}\frac{\vert \rho \vert^2}{\langle |\rho|^2 \rangle} \right\rangle \,.
\label{eq:beta:rho}
\eeqn

\begin{figure}
\begin{center}
\begin{tabular}{ll}
\includegraphics[width=53mm, angle=0]{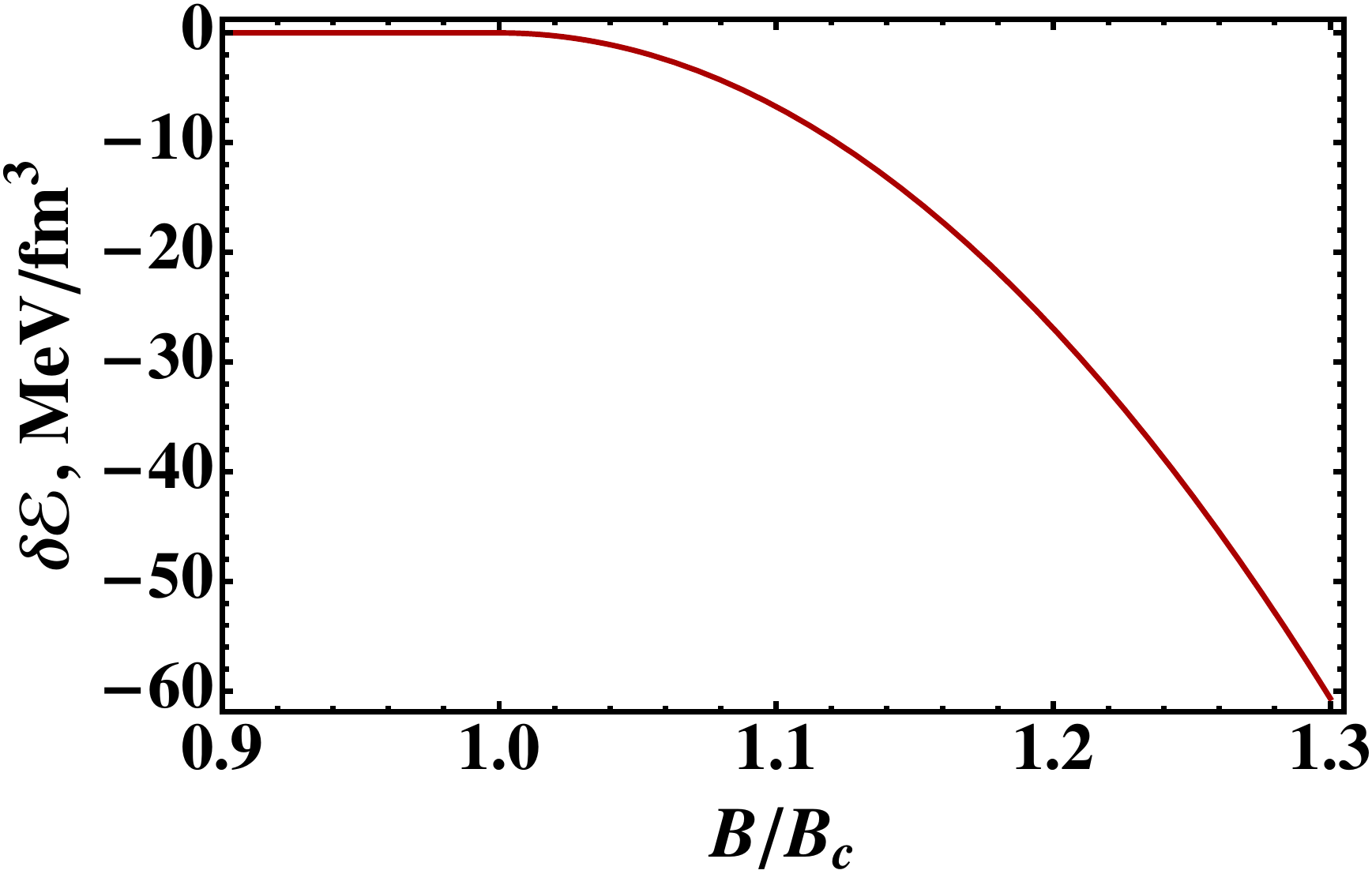}
&
\hskip 5mm \includegraphics[width=51mm, angle=0]{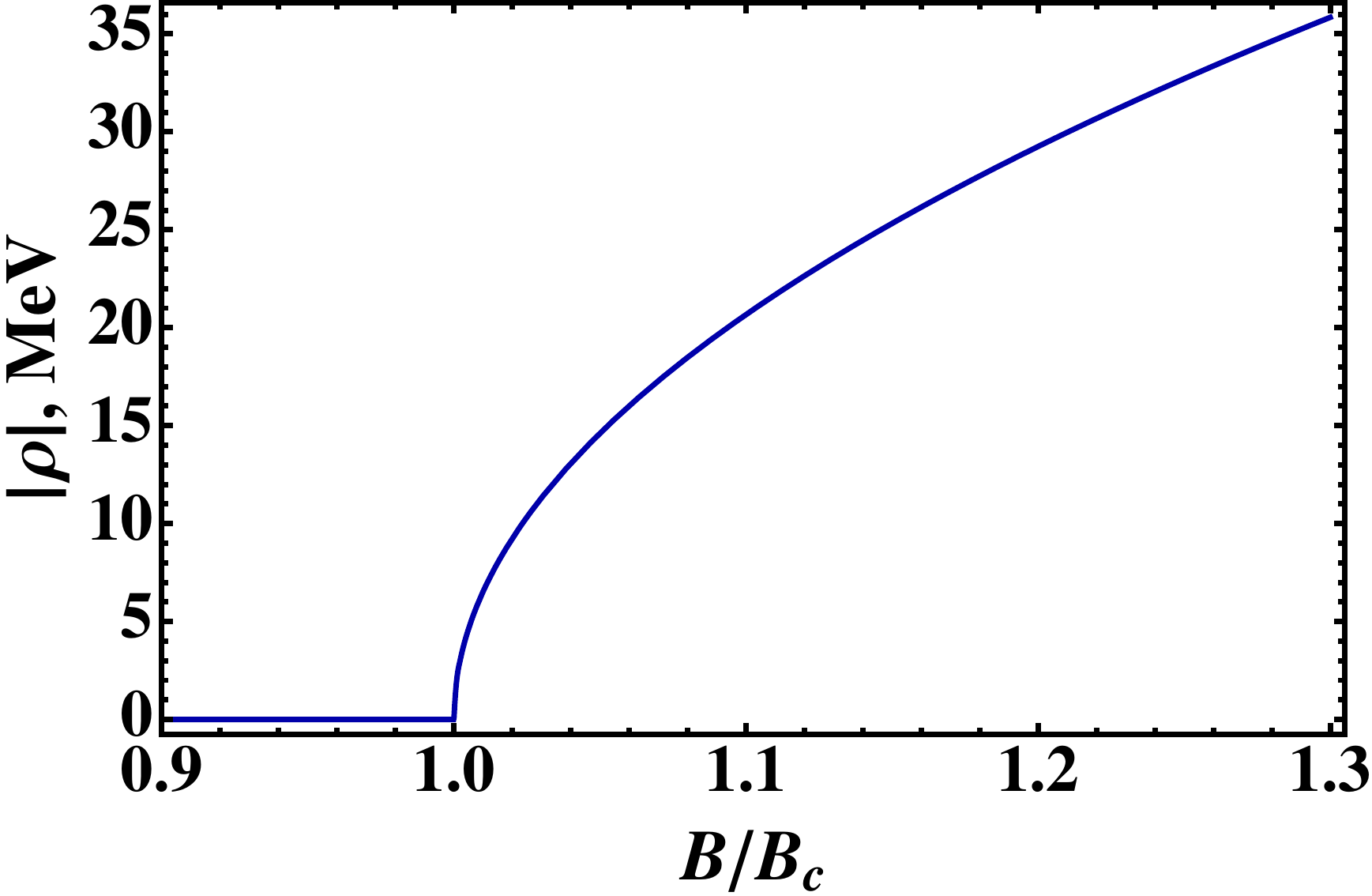}
\end{tabular}
\end{center}
\caption{At $B > B_c$ the superconducting state -- corresponding to the equilateral triangular lattice~Fig.~\ref{fig:GL:lattices}(right)  --  is more energetically favorable compared to the trivial vacuum state: at the critical magnetic field $B=B_c$, the condensation energy~\eq{eq:E:cond}, the left panel, becomes negative due to emergence of the superconducting condensate (the right panel). From Ref.~\cite{ref:Jos}.}
\label{fig:rho:rho}
\end{figure}

The left and right panels of Fig.~\ref{fig:rho:rho} show, respectively, the condensation energy 
\beqn
\delta {\cal E} = \langle {\cal E} \rangle - \frac{1}{2} B_\ext^2\,,
\label{eq:E:cond}
\eeqn
and the superconducting condensate $|\rho| \equiv \sqrt{\langle |\rho|^2 \rangle}$ in the ground state. The rise of the condensate at $B > B_{c}$ makes the ground state energy smaller compared to the normal, noncondensed state. 

\subsubsection{Periodic pattern of ground state: superfluid and superconductor vortices}
\label{sec:vortices}

Solutions of Eq.~\eq{eq:rho:z} are inhomogeneous functions in the transversal $(x_1,x_2)$ plane. The inhomogeneities in the charged $\rho$ condensate induce an unexpected condensation of the {\it neutral} $\rho$ mesons:
\beqn
\rho^{(0)}(x^\perp) 
= \frac{2 i g_s}{- \partial^2_{\perp} + m^2_0} \partial |\rho|^2
\equiv \frac{i g_s}{\pi} \partial \int d^{2} y^{\perp} K_0 \left(m_{0} |x^\perp - y^\perp|\right) |\rho (y^{\perp})|^2\,.
\label{eq:rho0:rho}
\eeqn
The longitudinal components of the neutral condensate are zero, $\rho^{(0)}_0 = \rho^{(0)}_3 = 0$.

The external magnetic field $B_{\ext}$ leads to a backreaction from the charged condensate \eq{eq:rho:z}, which creates a transverse electric current~\eq{eq:J:perp},
\beqn
J^{\perp}(x^{\perp}) = \Bigl(\frac{2 i e m_0^2  \partial}{- \partial_{\perp}^2 + m_0^2} |\rho|^2\Bigr)(x^\perp) 
\equiv \frac{i e m_{0}^{2}}{\pi} \partial \int d^{2} y^{\perp} K_0 \left(m_{0}|x^\perp - y^\perp|\right) |\rho (y^{\perp})|^2, \
\label{eq:j:explicit}
\eeqn
and, consequently, affects the magnetic field inside the superconductor:
\beqn
B(x^{\perp}) = B_\ext + \frac{2 e m_0^2}{- \partial_{\perp}^2 + m^2_0} \Bigl[|\rho(x^{\perp})|^2 - \langle{|\rho|^2}\rangle\Bigr] \,,
\qquad
\label{eq:B:solution}
\eeqn

\begin{figure}
\begin{center}
\begin{tabular}{ll}
\includegraphics[width=54mm, angle=0]{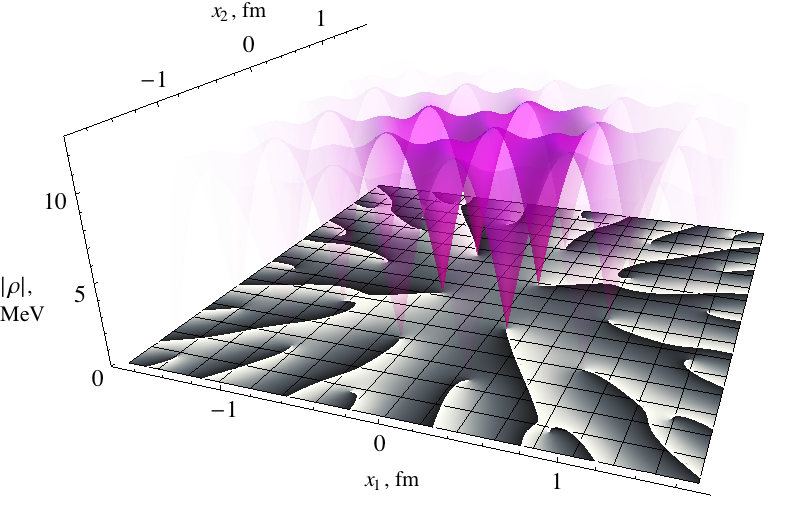} &
\hskip 3mm \includegraphics[width=56mm, angle=0]{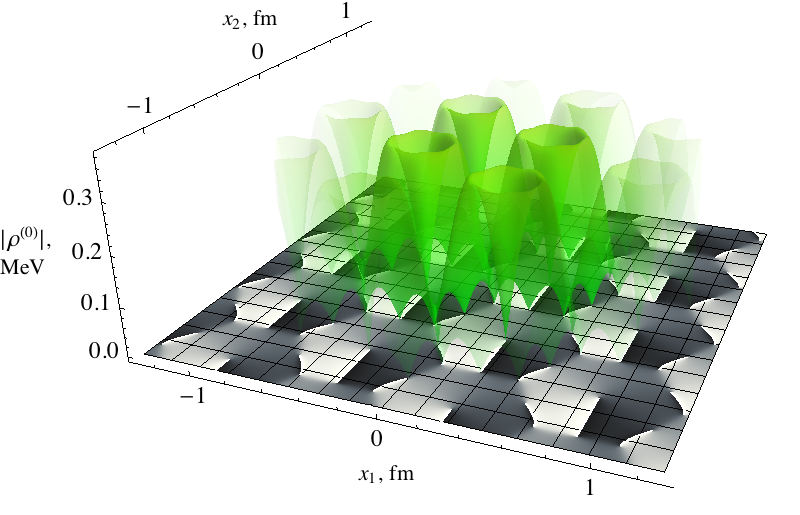}
\end{tabular}
\end{center}
\caption{The charged superconducting (left) and neutral superfluid (right) condensates in the transversal $(x_1,x_2)$ plane at  $B=1.01\,B_c$. The 3D plots illustrate the absolute values of the condensates (in MeV) while the corresponding projections on the $(x_{1},x_{2})$ planes of these figures are the density plots of the phases of the corresponding condensates. The white lines of the projections are gauge-dependent singularities (the Dirac sheets) which are (left) attached to the superconductor vortices and (right) stretched between the superfluid vortices and antivortices. }
\label{fig:absolute}
\end{figure}

\begin{figure}
\sidecaption
\includegraphics[width=45mm,angle=0]{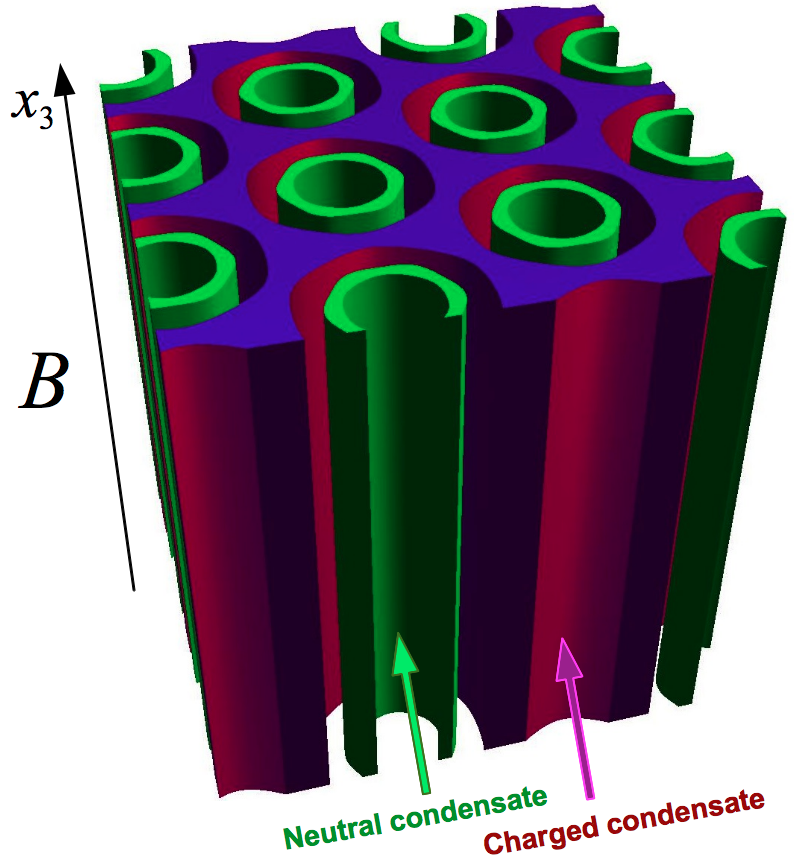}
\caption{A visualization of the nested structure of the electrically charged, superconducting condensate~\eq{eq:rho:z} and the electrically neutral, superfluid condensate~\eq{eq:rho0:rho}, plotted in magenta and green, respectively. Shown are the regions where these condensates take maximal values. The transverse plane corresponds to $2 \, \mathrm{fm} \times \, 2 \, \mathrm{fm}$ region at the magnetic field $B = 1.01 B_{c}$.}
\label{fig:nested}     
\end{figure}

In Fig.~\ref{fig:absolute} the charged and neutral condensates are plotted as functions of the transverse coordinates $x_{1}$ and $x_{2}$ for the magnetic field $B=1.01\,B_c$. The periodic equilateral-triangle structure of the absolute value of the charged $\rho$--meson condensate is identical -- apart from the magnitude of the physical scales -- to the one of the GL model, Fig.~\ref{fig:GL:lattices}(right). The absolute value of the neutral condensate also exhibits a lattice pattern which has, however, a bit more involved appearance: hexagonally-shaped structures are arranged into the equilateral triangular lattice, Fig.~\ref{fig:nested}.

The charged and the neutral $\rho$--meson condensates coexist together. Since the magnetic field cannot directly induce the neutral condensate, the mechanism of its appearance is as follows: the background magnetic field induces the charged condensate $\rho$, Eq.~\eq{eq:rho:z}, while the charged condensate gives rise to the neutral one, $\rho^{(0)}$, Eq.~\eq{eq:rho0:rho}. As a result, the neutral condensate is an order of magnitude smaller than the charged condensate, Fig.~\ref{fig:absolute}. These condensates form a nested structure, Fig.~\ref{fig:nested}.

The projection of Fig.~\ref{fig:absolute} (left) shows the density plot of the phase of the charged condensate, $\arg \rho$. The phase -- which is not a periodic function of the transverse coordinates $x_{1}$ and $x_{2}$ -- exhibits discontinuities across which the phase is changed by $2\pi$. These discontinuities correspond the Dirac sheets, shown as the white lines in the same projection of Fig.~\ref{fig:absolute} (left). The Dirac sheets are attached to the new class of vortices, ``the superconductor $\rho$ vortices''. The positions of these vortices correspond to the endpoints of the Dirac sheets. According to the 3D plot of the same figure, the absolute value of the $\rho$--meson condensate is vanishing at the centers of the superconductor vortices. Locally, these superconductor vortices have the structure which is similar (up to a gauge--dependent phase) to the Abrikosov vortices in the conventional superconductors~\eq{eq:Phi:vort}: $\rho (x^\perp)  \equiv \rho_{1} (x^\perp)  \equiv i \rho_{2} (x^\perp) \propto |x^\perp| e^{i \varphi} \equiv x_1 + i x_2$.

\begin{figure}
\begin{center}
\begin{tabular}{ll}
\includegraphics[width=50mm, angle=0]{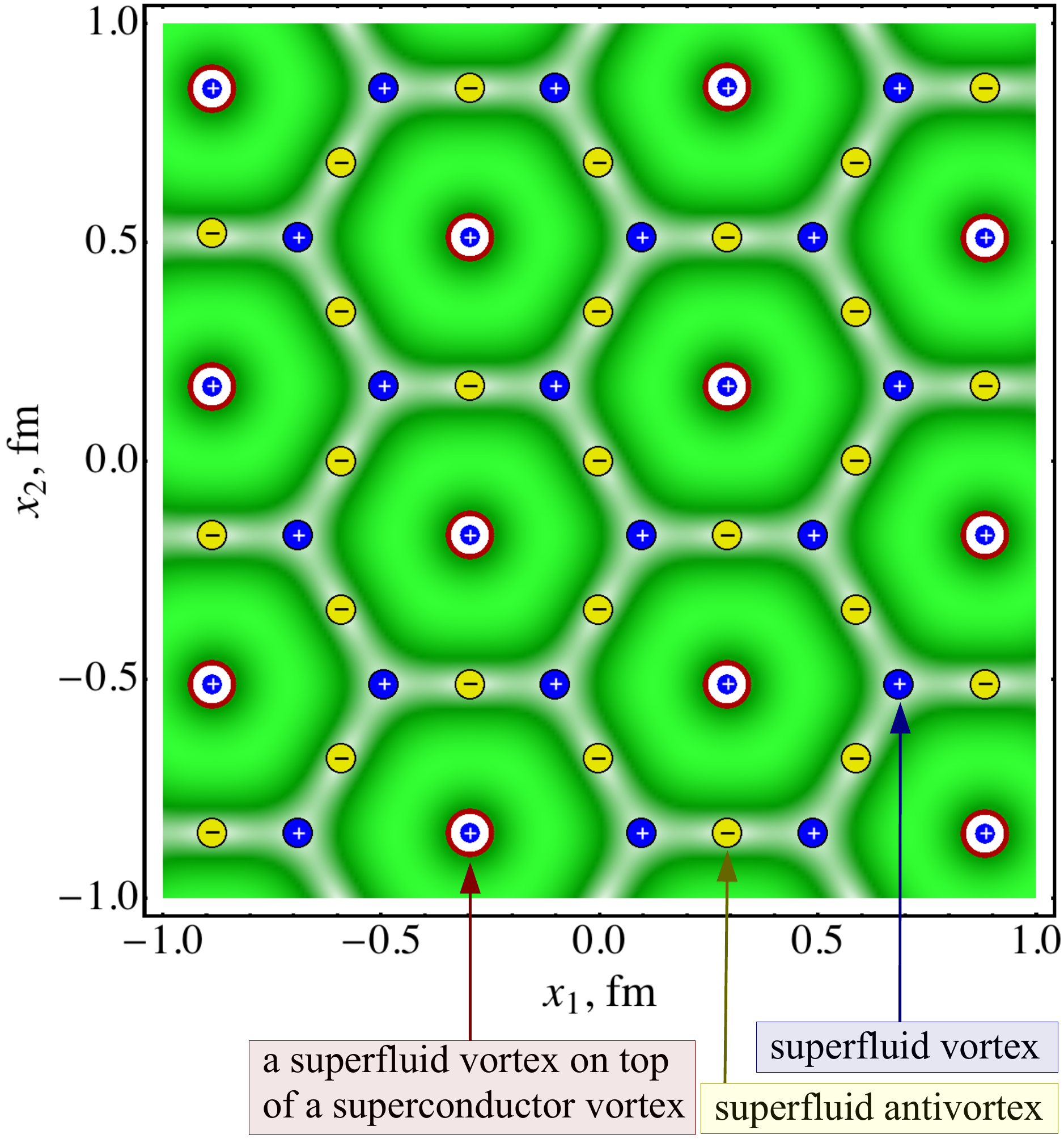} & \\[-54mm]
& \hskip 5mm \includegraphics[width=50mm, angle=0]{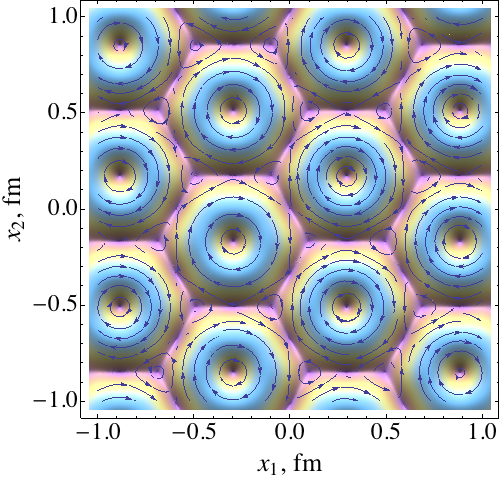} 
\end{tabular}
\end{center}
  \caption{(left) The kaleidoscopic vortex structure of the charged and neutral condensates induced by the magnetic field $B=1.01\,B_c$: the superconductor vortices (the large red circles) are always superimposed on the superfluid vortices (the small blue disks marked by the plus signs) forming an equilateral triangular lattice in the $(x_{1},x_{2})$ plane. The isolated superfluid vortices and antivortices (the small yellow disks with the minus signs) are arranged in the hexagonal lattice. The shades of green illustrate the absolute value of the neutral $\rho$--meson condensate~\eq{eq:rho0:rho} (from Ref.~\cite{ref:Jos}). (right) The density and the vector flow of the superconducting currents  in the $(x_{1},x_{2})$ plane at $B=1.01\,B_c$. }
\label{fig:rho:rho0}
\end{figure}

Contrary to the phase of the charged condensate~\eq{eq:rho:z}, the phase of the neutral condensate~\eq{eq:rho0:rho} is a periodic function of the $x_{1}$ and $x_{2}$ coordinates. The neutral phase exhibits the $2 \pi$--discontinuities as well. These discontinuities are visualized as white lines in the projection on the bottom-left panel of Fig.~\ref{fig:absolute}. The end-points of these discontinuities mark positions of a new type of vortices called ``superfluid $\rho$ vortices'' connected by the corresponding $2 \pi$--discontinuity to the superfluid antivortices. Locally, the superfluid vortices have, up to a phase, the familiar structure: $\rho^{(0)} (x^\perp) \equiv \rho^{(0)}_{1} (x^\perp) + i \rho^{(0)}_{2} (x^\perp)  \propto |x^\perp| e^{i \varphi} \equiv x_1 + i x_2$.

The vacuum ground state has a rich ``kaleidoscopic'' structure in terms of the vortex content: the equilateral triangular lattice of the superconductor vortices is superimposed on the hexagonal lattice of the superfluid vortices and antivortices. An elementary lattice cell of the kaleidoscopic lattice contains one superconductor vortex in the electrically charged $\rho$ condensate as well as three superfluid vortices and three superfluid antivortices in the neutral $\rho^{(0)}$ condensate, Fig.~\ref{fig:rho:rho0}(left).

 \subsubsection{Superconductivity and superfluidity in the ground state}

Electric transport properties of a material (such as, for example, the electrical conductivity) are usually determined in a linear response approximation in which one studies an electric current generated inside the material by a weak (test) external electric field background. The electric field should be weak enough in order to preserve, in a leading order, the ground state of the studied material.

In our ground state the transverse (with respect to the strong magnetic field) electric currents~\eq{eq:j:explicit} of charged condensates are confined to elementary cells of the periodic ground state, Fig.~\ref{fig:rho:rho0}(right). The size of the elementary cell is of the order of the size of the wavefunction of lowest Landau level\footnote{In physical units the size of the cell is approximately $0.5\,{\mathrm{fm}}$ for the near--critical field, $B \sim B_c$.}. In order for a net electric current to be induced in the transverse directions, the quarks need to be excited to the next Landau level which is, however, separated from the lowest Landau level by a large energy gap of the order of $\delta E \sim \sqrt{|e B|}$. It is the energy gap which makes the vacuum state to behave as an insulator in the transverse directions because a weak ($|\vec E| \ll |\vec B|$) transverse ($\vec E \perp \vec B$) electric field $\vec E$ cannot create large enough excitation of overcome the gap. The presence of the gap is the very reason why the Meissner effect is absent in the superconducting ground state~\cite{Chernodub:2010qx} so that the emerging superconductivity does not screen the external magnetic field (Section~\ref{sec:coexistence}, page~\pageref{sec:coexistence}).

\begin{figure}
\begin{center}
\begin{tabular}{ll}
\includegraphics[width=57mm, angle=0]{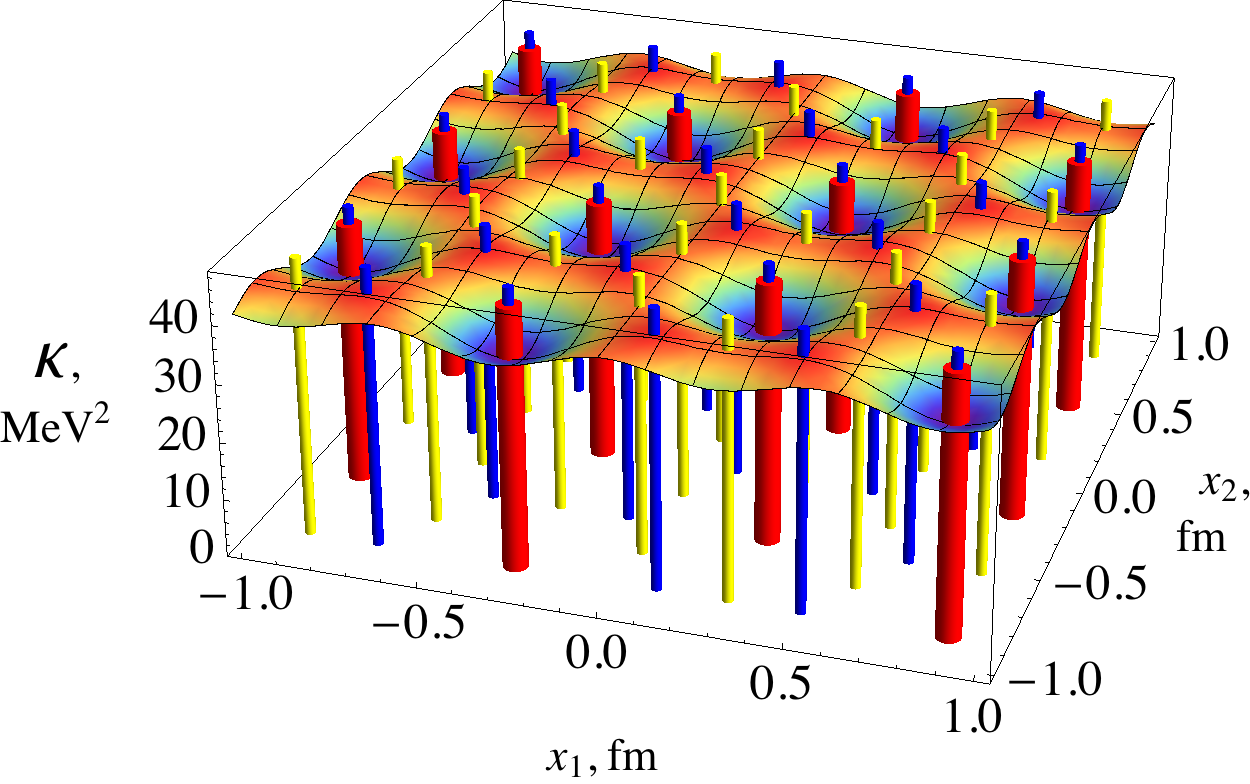} & 
\hskip 5mm \includegraphics[width=55mm, angle=0]{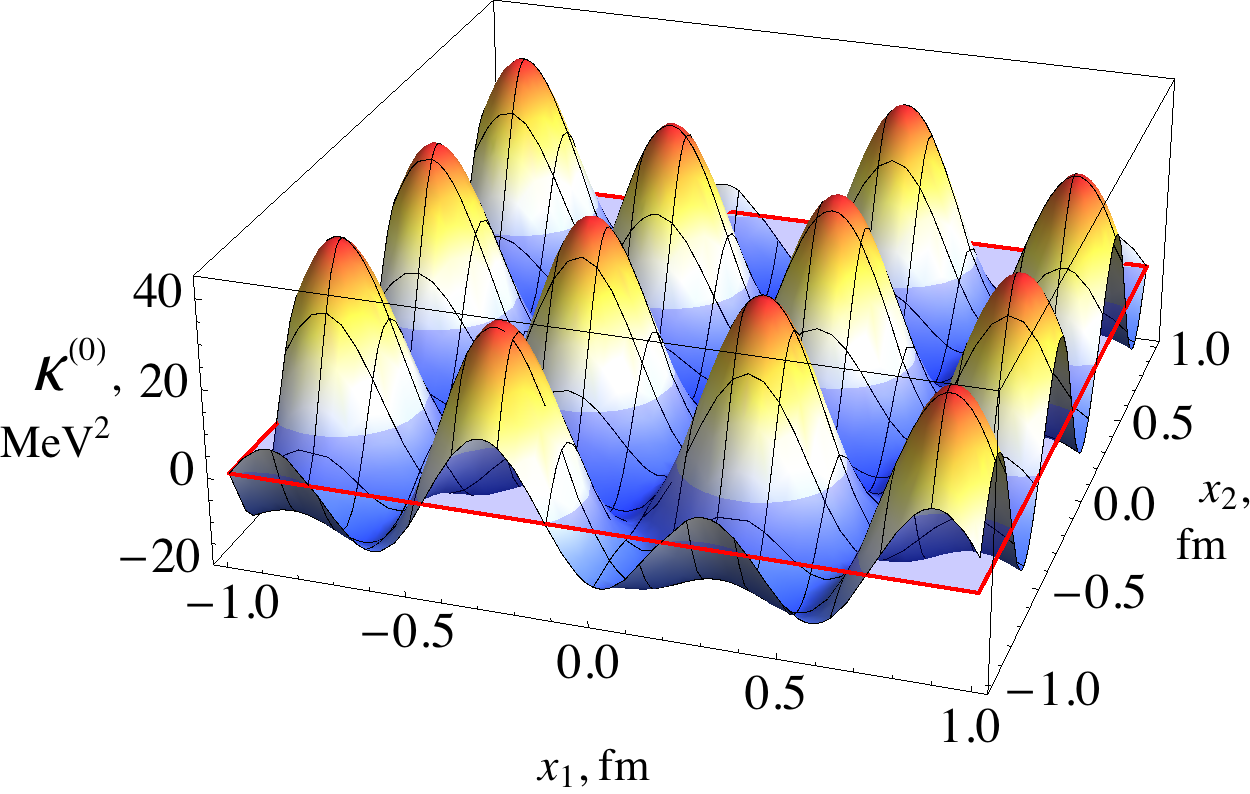} 
\end{tabular}
\end{center}
\caption{The strength of the vacuum (left) superconductivity $\kappa$, Eq.~\eq{eq:kappa}, and (right) superfluidity $\kappa^{(0)}$, Eq.~\eq{eq:kappa:rho0}, as the function of the transverse coordinates $x_{1}$ and $x_{2}$ at magnetic field $B=1.01\,B_c$. In addition, the left plot illustrates the superconductor vortices (the large red tubes) and the superfluid vortices and antivortices (the smaller blue and yellow tubes, simultaneously) in accordance with notations of Fig.~\ref{fig:rho:rho0}(left). In the right plot the semitransparent plane highlights the line $\kappa^{(0)} = 0$ where the superfluid strength changes its sign.}
\label{fig:currents}
\end{figure}

Contrary to the transverse electric currents, the longitudinal currents are not restricted by the external magnetic field. Let us apply a weak electric field $\vec E = (0,0,E_z)$ along the axis of the strong magnetic field $\vec B \equiv (0,0,B)$. According to the equations of motion of the $\rho$--meson model~\eq{eq:L}, the induced electric currents~\eq{eq:Jmu:2} satisfy the following equations~\cite{Chernodub:2010qx}:
\beqn
\frac{\partial J_{3}(x)}{\partial x^{{0}}}- \frac{\partial J_{{0}}(x)}{\partial x^{3}}  = - \kappa(x^\perp) E_z\,, 
\qquad \quad
\frac{\partial J_{{k}}(x)}{\partial x^{\mu}}- \frac{\partial J_{{\mu}}(x)}{\partial x^{k}}  \equiv 0 \,,
\label{eq:London}
\eeqn
where $\mu = 0,\dots,3$ and $k = 1,2$.

The set of equations~\eq{eq:London} is nothing but an anisotropic ``vacuum'' analogue of the London equation~\eq{eq:dJ:London} of superconductivity. Equations~\eq{eq:London} show that the electric current -- induced by a weak electric ``test'' field -- flows without resistance along the magnetic field axis while in the transverse directions the superconductivity is absent. The strength of the vacuum superconductivity is characterized by the quantity $\kappa$, which is a nonlocal function of the superconducting $\rho$--meson condensate:
\beqn
\kappa(x^\perp) = \Bigl( \frac{4 e^2 m_0^2}{- \partial_{\perp}^2 + m_0^2}  |\rho|^2\Bigr)(x^\perp)
\equiv \frac{2 e^2 m_0^2}{\pi} \int d^{2} y^{\perp} K_0 \left(m_{0} |x^\perp - y^\perp|\right) |\rho (y^{\perp})|^2. \
\label{eq:kappa}
\eeqn

According to Fig.~\ref{fig:currents}(left), the strength of the superconductivity~\eq{eq:kappa} is a weakly dependent function of the transverse coordinates $x_{1}$ and $x_{2}$. In a response to a weak electric current, the superconducting currents are generated outside of the superconductor vortex cores and the maxima of the induced electric currents are located at the centers of the superfluid vortices. Contrary to the ordinary superconductivity, the vacuum superconductivity is not completely suppressed inside the vortices due to the nonlocal nature of the relation between the transport coefficient $\kappa$ and the superconducting condensate~\eq{eq:kappa}.

Unexpectedly, the condensate of the neutral $\rho$ mesons is also sensitive to the presence of the external electric current. It turns out that the electrically neutral current of the $\rho^{(0)}$ mesons, defined via the relation $J_{\mu}^{(0)} = - \frac{e}{g_s} \partial^\nu f^{(0)}_{\nu\mu}$, satisfies a London--like equation as well~\cite{Chernodub:2012bj}:
\beqn
\frac{\partial J^{(0)}_{3}(x)}{\partial x^{0}}- \frac{\partial J^{(0)}_{0}(x)}{\partial x^{3}} & = & - \kappa^{(0)} (x_{1},x_{2}) E_z (x)\,,
\label{eq:London:rho0} \\
\kappa^{(0)}(x_{1},x_{2}) & = & \Bigl( \frac{4 e^2 \partial_{\perp}^2}{ - \partial_{\perp}^2 + m_0^2}  |\rho|^2\Bigr)(x_{1},x_{2}) \equiv \frac{1}{m_{0}^{2}} \partial^{2}_{{\perp}} \kappa(x^\perp)\,,
\quad
\label{eq:kappa:rho0}
\eeqn
where the superfluid coefficient $\kappa^{(0)}$ is visualized in Fig.~\ref{fig:currents}(right).

Equations \eq{eq:London} and \eq{eq:London:rho0} indicate that, respectively, the charged and neutral currents should flow frictionlessly (i.e., accelerating ballistically) along the magnetic field axis if an weak external electric field is applied along the same direction. Then, if even at some moment of time the electric field is set back to zero, both the superconducting current and the superfluid flow would continue to flow permanently because of the absence of the dissipation forces for the corresponding condensates. 

Notice that the electric--field--induced superfluid flow is a locally nonvanishing quantity, while the total superfluid flow of each elementary lattice cell is zero,
\beqn
\frac{\partial \langle J^{(0)}_{3} \rangle_{{\perp}}(x)}{\partial x^{0}} - \frac{\partial \langle J^{(0)}_{0} \rangle_{{\perp}}(x)}{\partial x^{3}} = 0\,, 
\qquad \quad 
\langle \cO \rangle_{\perp} \equiv \int d^{2} x^{\perp} \cO(x)\,,
\label{eq:average}
\eeqn
because the cell--averaged superfluid coefficient~\eq{eq:kappa:rho0} is zero, $\langle \kappa^{(0)}(x^{\perp}) \rangle_{\perp} = 0$, too.
A comparison of the left and right plots of Fig.~\ref{fig:currents} reveals that the external electric field generates the positive superfluid flow at the positions of the superconductor vortices which are always accompanied by a superfluid vortices according to Fig.~\ref{fig:rho:rho0}(left). The negative superfluid flow is generated at the positions of other, unaccompanied superfluid vortices and antivortices.

Finally, we would like to stress that the global quantum numbers of the new exotic superconducting (and, simultaneously, superfluid) phase correspond to the quantum numbers of vacuum. For example, all chemical potentials in the superconducting phase are vanishing. The vacuum is an electrically neutral object: the presence of the positively charged condensate $\rho$  implies an automatic appearance of a negatively charged condensate $\rho^*$ of the equal magnitude, $\rho \equiv |\rho^*|$.  As a result, in strong magnetic field the energy of the vacuum is lowered due to the emergence of the charged condensates, while the net electric charge of the vacuum stays always zero~\cite{Chernodub:2010qx,Chernodub:2011mc}. Despite of the net electric neutrality, the vacuum exhibits the superconductivity since a weak external electric field  pushes the positively and negatively charged condensates in opposite directions along the magnetic field axis, thus creating a net electric current of a double magnitude.

\subsubsection{Abelian gauge symmetry breaking and gauge--Lorentz locking}

What is the symmetry breaking pattern in the new superconducting phase of the vacuum? The vacuum superconductivity appears due to the emergence of the magnetic--field--induced $\rho$--meson condensate~\eq{eq:ground:state}. In the presence of the background magnetic field ${\vec B}$, the group of global rotations $SO(3)_{\mathrm{rot}}$ of the coordinate space is explicitly broken to its $O(2)_{\mathrm{rot}}$ subgroup which is generated by rotations around the magnetic field axis. The scalar combination $\rho$ of the vector condensates~\eq{eq:ground:state} transforms under the residual rotational group $O(2)_{\mathrm{rot}}$ as follows:
\beqn
O(2)_{\mathrm{rot}}: \qquad \rho(x) & \to & e^{i \varphi} \rho(x)\,,
\label{eq:rot:invariance:rho}
\eeqn
where $\varphi$ is the azimuthal angle of the rotation in the transverse plane about the $x_{3}$ axis. The $\rho$ meson field transforms also under the electromagnetic group~\eq{eq:gauge:transformations}:
\beqn
U(1)_{\mathrm{e.m.}}: \qquad \rho(x) & \to & e^{i e \omega(x)} \rho(x)\,,
\label{eq:gauge:invariance:rho}
\eeqn
Thus, if the condensate $\rho$ were a homogeneous (i.e., coordinate independent) quantity then the ground state would ``lock'' the electromagnetic gauge symmetry with the residual rotational symmetry\footnote{A philosophically similar phenomenon, a color-flavor locking, is realized in a different context of the color superconductivity in a dense quark matter~\cite{Alford:2001dt}.}, $U(1)_{\mathrm{e.m.}} \times O(2)_{\mathrm{rot}} \to U(1)_{\mathrm{locked}}$, since a rotation of the coordinate space at the angle $\varphi$ about the axis $x_{3}$ and a simultaneous gauge transformation with a constant gauge-angle $\omega = - \varphi/e$ leave the homogeneous condensate $\rho$ intact. The inhomogeneities in the $\rho$ condensate break the locked subgroup further from the global $U(1)$ group down to a discrete subgroup of the lattice rotations $G^{\mathrm{lat}}_{\mathrm{locked}}$ of the kaleidoscopic lattice state, Fig.~\ref{fig:rho:rho0}(left), Ref.~\cite{Chernodub:2010qx}:
\beqn
U(1)_{\mathrm{e.m.}} \times O(2)_{\mathrm{rot}} \to G_{\mathrm{locked}}^{\mathrm{lat}}\,.
\label{eq:locking2}
\eeqn

\subsection{Superconductivity of vacuum in Nambu--Jona-Lasinio model}
\label{sec:NJL}

Basic properties of ordinary superconductivity can equally be revealed either in the Ginzburg--Landau (GL) phenomenological approach which describes the scalar field of the superconducting carriers or in the Bardeen--Cooper-Schrieffer (BCS) model which accounts the electron pairing into the superconducting carriers (Section~\ref{sec:GL:BCS}, page~\pageref{sec:GL:BCS}). Both approaches are mathematically equal if the temperature is sufficiently close to the superconducting phase transition~\cite{ref:equivalence}.

So far we have discussed the basic features of the vacuum superconductivity in the effective $\rho$ meson electrodynamics~\cite{Djukanovic:2005ag}, which serves as a ``vacuum'' analogue of the GL approach to the ordinary superconductivity~\cite{ref:GL}. Below we briefly consider, following Ref.~\cite{Chernodub:2011mc}, the $\rho$--meson condensation in the BCS--like approach~\cite{ref:BCS}, which is based on the Nambu-Jona--Lasinio model~\cite{ref:NJL,Ebert:1985kz}.

\subsubsection{Effective action in strong magnetic field}

We consider an extended two-flavor ($N_f = 2$) three--color ($N_c = 3$) Nambu-Jona-Lasinio (NJL) model~\cite{Ebert:1985kz}:
\beqn
\cL(\psi,\bar\psi)  & = & \bar \psi \bigl(i \slashed \partial + {\hat Q} \, {\slashed {\cal A}} - \hat M^0 \bigr) \psi 
+ \frac{G^{(0)}_S}{2} \bigl[\bigl(\bar \psi \psi\bigr)^2  + \bigl(\bar \psi i\gamma^5 \vec \tau \psi\bigr)^2 \bigl] \nonumber \\
& & - \frac{G^{(0)}_V}{2}  \sum\nolimits_{i=0}^{3} \left[\bigl(\bar \psi \gamma_\mu \tau^i\psi\bigr)^2 + \bigl(\bar \psi  \gamma_\mu \gamma_5  \tau^i \psi\bigr)^2\right]\,,
\label{eq:L:NJL}
\eeqn
where the light quarks are represented by the doublet $\psi = (u,d)^T$, and $G^{(0)}_S$ and $G^{(0)}_V$ are the corresponding bare couplings of scalar and vector quarks' interactions. The masses $m_{u}$ and $m_{d}$, and electric charges ($q_u = +2e/3$ and $q_d = -e/3$) of the quarks are combined into the bare mass matrix $\hat M^0 = {\mathrm{diag}} (m^0_u, m^0_d)$ and the charge matrix $\hat Q = {\mathrm{diag}} (q_u,q_d)$, respectively. The $2 \times 2$ matrices in the flavor space are denoted by hats over the corresponding symbols and $\tau^{i}$, $i=1,2,3$, are the Pauli matrices.

The partition function of the NJL model can be represented as an integral,
\beqn
\cZ = \int D \bar \psi  D \psi \ e^{i \int d^4 x \, \cL} = \int D \sigma D \pi D V D A \, e^{i S[\sigma,\vec\pi,V,A]}\,,\quad
\label{eq:Z:NJL}
\eeqn
over bosonic fields~\cite{Ebert:1985kz}.  The bosonic fields are given by one scalar field $\sigma \sim \bar\psi \psi$, the triplet of three pseudoscalar fields $\vec \pi \sim \bar\psi \gamma^5 \vec \tau \psi$ [made of the electrically neutral, $\pi^0 \equiv \pi^3$, and electrically charged, $\pi^\pm = (\pi^{1}\mp i \pi^{2})/\sqrt{2}$, pions], four vector fields,
\beqn
{\hat V}_\mu & \equiv & \sum\nolimits_{i=0}^3 \tau^i V_\mu^i = \matrix{\omega_\mu + \rho^{0}_\mu}{\sqrt{2} \rho^+_\mu}{\sqrt{2} \rho^-_\mu}{\omega_\mu - \rho^{0}_\mu}\,,
\qquad
V_\mu^i \sim \bar\psi \gamma_\mu \tau^i \psi\,,
\label{eq:matrix:U}
\eeqn
[composed of the flavor-singlet coordinate-vector $\omega$--meson field $\omega_\mu$, and of the electrically neutral, $\rho^0_\mu \equiv \rho^3_\mu$, and charged, $\rho^\pm_\mu = (\rho^{1}_\mu \mp i \rho^{2}_\mu)/\sqrt{2}$, components of the $\rho$-meson triplet], and four pseudovector (axial) fields,
\beqn
{\hat A}_\mu & \equiv & \sum\nolimits_{i=0}^3 \tau^i A_\mu^i = \matrix{f_\mu + a^{0}_\mu}{\sqrt{2} a^+_\mu}{\sqrt{2} a^-_\mu}{f_\mu - a^{0}_\mu}\,, 
\qquad
A_\mu^i\sim \bar\psi \gamma^5 \gamma_\mu \tau^i \psi\,.
\label{eq:matrix:A}
\eeqn
where the fields $f_\mu$ and $(a^0_\mu,a^\pm_\mu)$ represent the singlet axial $f_1$ meson and the $\vec a_1$ triplet of the axial mesons, respectively.

The effective bosonic action in Eq.~\eq{eq:Z:NJL} is as follows
\beqn
& & S[\sigma,\vec\pi,V,A] = S_\psi - \int d^4 x \Bigl[\frac{1}{2 G^{(0)}_S} (\sigma^2 + \vec \pi^2) - \frac{1}{2 G^{(0)}_V} (V^k_\mu V^{k\mu} + A^k_\mu A^{k\mu}) \Bigr],
\label{eq:S}\\
& & S_\psi[\sigma,\vec\pi,V,A] = - i N_c {\mathrm {Tr}}\,  {\mathrm{Ln}}(i \cD)\,,
\label{eq:S:psi} \\
&  & \, i \cD = i {\slashed \partial} + {\hat Q} \, {\slashed {\cal A}} - \hat M^0 + {\hat {\slashed V}}_\mu + \gamma^5 {\hat {\slashed A}} - (\sigma + i \gamma^5 \vec \pi \vec \tau)\,. \quad
\label{eq:icD}
\eeqn
where we have used simplified notations for the expectation values of the fields, $\langle\sigma\rangle = \sigma$ etc. In the absence of a magnetic field background the expectation value of $\sigma$ plays a role of the constituent quark mass, $m_q = \sigma \sim 300$~MeV while the expectation values of the fields $\vec\pi$, $V$, and $A$ are zero~\cite{Ebert:1985kz}.

The effective action~\eq{eq:S} in the strong magnetic field background was calculated in Ref.~\cite{Chernodub:2011mc} in the lowest Landau level (LLL) approach using a mean--field technique inspired by calculations of the magnetic catalysis phenomenon~\cite{ref:Igor:review}. In the regime of the LLL dominance the propagator of a $f$'s quark
\beqn
S_f^{\mathrm{LLL}}(x,y) = P^\perp_f(x^{\perp},y^{\perp}) \, S^\parallel_f(x^{\parallel} - y^{\parallel})\,,
\label{eq:S:f}
\eeqn
factorizes into the $B$--transverse projector onto the LLL states
\beqn
P^\perp_f(x^{\perp},y^{\perp}) = \frac{|q_f B|}{2 \pi}  e^{\frac{i}{2} q_f B \varepsilon_{ab} x^{a} x^{b} - \frac{1}{4} |q_f B| (x^\perp - y^\perp)^{2}}\,, \qquad
\label{eq:P:perp}
\eeqn
and $B$-longitudinal fermion pro\-pa\-ga\-tor in the $1+1$ dimensions,
\beqn
S^\parallel_f(k_\parallel) \equiv S^\parallel_{\mathrm{sign}} (k_\parallel) = \frac{i}{\gamma^\parallel k_\parallel - m} P^\parallel_f \,,
\qquad
P^\parallel_f = \frac{\bbbone - i f \gamma^1 \gamma^2}{2}\,,
\label{eq:Sf:0}
\eeqn
which depend, respectively, on the $B$-transverse, $x^{\perp} = (x^{1},x^{2})$, and $B$-longitudinal, $x^{\parallel} {=} (x^{0},x^{3})$, coordinates~\cite{ref:Igor:review}.
Here $q_f$ is the electric charge of the $f^{\mathrm{th}}$ quark ($f=u,d$) and $eB>0$ is taken for definiteness.  

In Equation~\eq{eq:Sf:0} the matrix $P^\parallel_f$ is the spin projector operator onto the fermion  states with the spin polarized along (for $u$ quarks) or opposite (for $d$ quarks) to the magnetic field (we use $f=\pm1$ for, respectively, $f=u,d$). The operator $P^\parallel_f$ projects the original four 3+1 fermionic states onto two (1+1)--dimensional fermionic states, so that in the LLL approximation the quarks can move only along the axis of the magnetic field (the latter fact is a natural sequence of the LLL dominance~\cite{ref:Igor:review}). 

The operator~\eq{eq:P:perp} satisfies the projector relation,
\beqn
P^\perp_f \circ P^\perp_f = P^\perp_f\,, 
\qquad
A \circ B \equiv {\int} d^2 y^\perp \, A(\dots,y^{\perp}) \, B(y^\perp,\dots)\,,
\eeqn
where  "$\circ$"  is  the convolution operator in the $B$-transverse space.

In the one-loop order the effective action \eq{eq:S:psi} contains a scalar and vector parts:
\beqn
S = S_{S} (\sigma,\vec \pi)+ S_{V} (A,V)\,,
\eeqn
respectively. In terms of the nontrivial condensates\footnote{Here we omit all terms with vanishing condensates as well as all kinetic terms.}, 
the potential term in the scalar part of the action has the following (renormalized) form:
\beqn
S_{S} = - \int d^4 x \, \left[\frac{1}{2 G_S} \sigma^2  + \frac{|eB| N_c}{8 \pi^2} \Bigl(\ln \frac{\sigma^2}{\mu^2}-1\Bigr)  \sigma^2 \right]\,,
\label{eq:S:S}
\eeqn
which reflects one of the most important features of the magnetic catalysis~\cite{ref:Igor:review}: an enhancement of quarks' masses by the magnetic field background,
\beqn
m_q(B) = \sigma_{\mathrm{min}}(B) = \mu \exp\{-  2 \pi^2/(G_S N_c |eB|)\}\,,
\label{eq:mq:B}
\eeqn
given by the minimum $\sigma_{\mathrm{min}}$ of potential~\eq{eq:S:S}. The mass scale, $\mu \propto \sqrt{|eB|}$ is to be fixed beyond the LLL approximation because it is determined, in particular, by the $(1+1)$ dimensional motion of the quarks along the magnetic field~\cite{ref:Igor:review}.  As noticed in Ref.~\cite{ref:Igor:review}, the renormalization of the scalar NJL coupling $G_{S}$ in the
$\overline{\mathrm{MS}}$ scheme, 
\beqn
\frac{1}{G_S} = \frac{1}{G^{(0)}_S} - \frac{N_c |eB|}{4 \pi^2 \oeps}\,, 
\qquad 
G_S  \equiv \frac{2 \pi G_{\GN}}{N_c |eB|}\,,
\eeqn
is very similar to the renormalization of the coupling constant $G_{\GN}$ in the 1+1 dimensional Gross-Neveu model~\cite{ref:GN}. The divergencies of the $1+1$ dimensional fermions are treated in the dimensional regularization in $d=2 - 2 \epsilon$ dimensions, $1/\oeps = 1/\epsilon - \gamma_E + \log 4 \pi$ and $\gamma_E \approx 0.57722$ is Euler's constant.

A potentially nontrivial part of the (non-renormalized yet) effective vector action,
\beqn
S^{(0)}_{V} & \equiv & \frac{i N_c}{2} {\mathrm {Tr}}\, \left[ \frac{1}{i \cD_0} ( {\hat {\slashed V}}_\mu + \gamma^5 {\hat {\slashed A}}) \frac{1}{i \cD_0} 
( {\hat {\slashed V}}_\mu + \gamma^5 {\hat {\slashed A}}) \right]  =  \frac{4 N_c |e B|}{ 9 \pi^2} 
\label{eq:S0:V}\\
& & \cdot \int d^2 x^{\parallel} \Bigl[\Bigl(\frac{1}{\overline{\epsilon}} - \ln\frac{\sigma^2}{\mu^2} \Bigr)  (\phi^* \circ P_e \circ \phi) 
+ \Bigl(\frac{1}{\overline{\epsilon}} - \ln\frac{\sigma^2}{\mu^2} - 2 \Bigr)  (\xi^* \circ P_e \circ  \xi)\Bigr],
\nonumber
\eeqn
involves only the $B$-transverse combinations of the vector and axial mesons,  $\phi = (\rho^+_1 + i \rho^+_2)/2$ and $\xi = (a^+_1 + i a^+_2)/2$. In Eq.~\eq{eq:S0:V} the $B$-transverse projector for the unit charged particle $P_{e}$ is given by Eq.~\eq{eq:P:perp} with the replacement $q_f \to e$:
\beqn
P^\perp_e(x^{\perp},y^{\perp}) = \frac{9 \pi}{|e B|} P^\perp_u(x^{\perp},y^{\perp})  P^\perp_d(y^{\perp},x^{\perp})\,.
\label{eq:P:e}
\eeqn

The potential~\eq{eq:S0:V} has an unstable tachyonic mode which is determined by an inhomogeneous eigenstate of the charge-1 projection operator~\eq{eq:P:e}:
\beqn
(P_e \circ \phi)(x^{\perp}) = \phi(x^\perp)\,.
\label{eq:Pe:eigenstate}
\eeqn
The solution to this equation is a general periodic Abrikosov-like configuration~\cite{ref:Abrikosov} which is given, up to a proportionality coefficient, by Eq.~\eq{eq:rho:z}: $\phi(x^{\perp}) \propto \rho(x^{\perp})$. One can also show~\cite{Chernodub:2011mc} that there are no unstable modes for the axial mesons and, in accordance with Eq.~\eq{eq:ground:state}, no unstable modes exist for $B$-longitudinal components of the $\rho$ mesons. Thus, we set the corresponding expectation values to zero.

For the sake of simplicity, we set in Eq.~\eq{eq:rho:z} all coefficients $C_{n}$ equal, $C_{n} = \phi_0$, and then we evaluate certain basic quantities for the simplest square lattice~\eq{eq:Abrikosov:lattice}. As we have mentioned, despite different visual appearances of the square and equilateral triangular lattices, Fig.~\ref{fig:GL:lattices}, the corresponding bulk quantities (as, for example, the energy density) evaluated at these condensate solutions are almost the same as the difference between them lies within (sometimes, a fraction of) percents. 

The leading quadratic and quartic terms in the effective potential for the square lattice solution~\eq{eq:Abrikosov:lattice} of the eigenvalue equation~\eq{eq:Pe:eigenstate} are given by 
\beqn
V {=} \sqrt{2} \left(\frac{1}{G_B} - \frac{2 N_c |eB|}{9 \pi^2} \right) |\phi_0|^2  + C_0 \frac{|eB| N_c}{2 \pi^2 m^2} |\phi_0|^4\,, 
\qquad
\frac{1}{G_B} = \frac{1}{G_V} - \frac{8}{9 G_S}\,,
\label{eq:V:24}
\eeqn
where $C_0 \approx 1.2$ is a numerical (geometrical) factor. If the magnetic field exceeds certain critical strength\footnote{We have estimated the critical field only approximately since the phenomenological values of the NJL parameters $G_{S,V}$ are not known precisely~\cite{ref:physical}. Moreover, subtleties of the renormalization of the effective dimensionally reduced $(1+1)$-dimensional theory embedded in 3+1 dimensions provide us with an additional uncertainty.}, $e B_c^\NJL  = 9 \pi^2/(2 N_c \, G_B) \sim 1\,{\mbox{GeV}}^2$, then the potential~\eq{eq:V:24} becomes unstable towards a spontaneous creation of the $B$-transverse $\rho$--meson condensates  with the tachyonic mode $\rho^-_1(x^{\perp}) = - i \rho^-_2(x^{\perp}) = \phi(x^{\perp})$ [and, respectively, $\rho^+_1(x^{\perp}) = i \rho^+_2(x^{\perp}) = \phi^{*}(x^{\perp})$], where $\phi(x^{\perp})$ is a solution of Eq.~\eq{eq:Pe:eigenstate}.

\subsubsection{Electromagnetically superconducting ground state in the NJL model}

In the magnetic field background, the effective $\rho$--meson potential in the NJL model~\eq{eq:V:24} has the same Ginzburg--Landau form as the potential~\eq{eq:epsilon:0:2} for the $\rho$--meson field in the $\rho$--meson electrodynamics~\eq{eq:L}. If the magnetic field exceeds the critical value, $B \geqslant B_c^\NJL$, then the charged $\rho$--meson condensate emerge. In terms of the quark fields the new vector quark--antiquark condensates are as follows:
\beqn
\langle \bar u \gamma_1 d\rangle = - i \langle \bar u \gamma_2 d \rangle = \frac{\phi_0(B)}{G_{V}} \, K\Bigl(\frac{x_1+ i x_2}{L_B}\Bigr)\,, \quad
\quad
\label{eq:ud:cond}
\eeqn
where the function $K(z)$ is given in Eq.~\eq{eq:K}. The magnitude and global phase $\theta_0$ of the condensate~\eq{eq:ud:cond} are determined by the following formula:
\beqn
\phi_0(B) = e^{i \theta_0} C_\phi m_q(B) {\left(1 - \frac{B_c^{\NJL}}{B}\right)}^{1/2} \qquad \mbox{for} \quad B > B_c^{\NJL}\,.
\label{eq:phi0}
\eeqn
Here $C_\phi \approx 0.51$ is a numerical prefactor and the $B$--dependent quark mass $m_q$ is given in Eq.~\eq{eq:mq:B}. At $B < B_c^\NJL$ the $\rho$--meson condensate~\eq{eq:phi0} is zero. The superconducting phase transition at $B=B_c$ is of the second order with the critical exponent~$1/2$, similarly to the phase transition in the $\rho$--meson electrodynamics~\eq{eq:L}.

The quark condensates~\eq{eq:ud:cond} have the quantum numbers of the $\rho$ mesons. They form an inhomogeneous ground state identical to the one found in the $\rho$--meson electrodynamics~\eq{eq:rho:z}. It is very interesting to notice that the ground state in the NJL model~\eq{eq:L:NJL}, determined by the integral equation~\eq{eq:Pe:eigenstate}, and the ground state in the $\rho$--meson electrodynamics~\eq{eq:L}, determined by the differential equation of motion~\eq{eq:cDrho0}, have exactly the same functional form~\eq{eq:rho:z}.

The vacuum state~\eq{eq:ud:cond} of the NJL model is superconducting. The validity of the anisotropic London equation~\eq{eq:London} for the quarks' electric current,
\beqn
J^\mu(x) = \sum\nolimits_{f=u,d} q_f  \langle \bar \psi_f \gamma^\mu \psi_f\rangle \equiv - \tr [\gamma^\mu {\hat Q} S(x,x)]\,, \qquad
\label{eq:Jmu:general}
\eeqn
can be shown in a linear-response approach using retarded Green functions~\cite{Chernodub:2011mc}:
\beqn
\frac{\partial \langle J_{{3}}\rangle (x^\parallel)}{\partial x^{0}} - \frac{\partial \langle J_{{0}}\rangle (x^\parallel)}{\partial x^{3}} = - \frac{2 C_q}{(2 \pi)^3} e^3 \bigl(B - B_c^{\NJL}\bigr) \, E_3 \qquad \mbox{for} \quad B > B_c^{\NJL}\,, \qquad
\label{eq:London:local}
\eeqn
where $C_q \approx 1$ is a numerical prefactor and at $B < B_c$ the right hand side of Eq.~\eq{eq:London:local} is zero. For the sake of simplicity, in Eq.~\eq{eq:London:local} we have averaged the electric charge density $J^0$ and the electric current $J^z \equiv J^{0}$ over the transverse plane~\eq{eq:average}. 

Apart from the prefactors, the London equations in the NJL model~\eq{eq:London:local} and in the $\rho$-meson electrodynamics~\eq{eq:London} are identical. In a linear--response approximation these laws can be generalized to a completely Lorentz-covariant form~\cite{Chernodub:2011mc}, 
\beqn
\partial_{\mu}  J_{{\nu}} - \partial_{\nu}  J_{{\mu}}  = \gamma \cdot  (F, {\widetilde F}) {\widetilde F}_{{\mu\nu}}\,, 
\eeqn
via the Lorentz invariants $(F, {\widetilde F}) =  4 (\vec B, \vec E)$ and $(F, F) = 2 (\vec B^2 - \vec E^2)$. Here $\gamma$ is a function of the scalar invariant $(F, F)$ and ${\widetilde F}_{\mu\nu} = \epsilon_{\mu\nu\alpha\beta} F^{\alpha\beta}/2$. The Lorentz-covariant forms of the local London laws for the superconductor~\eq{eq:London} and superfluid~\eq{eq:London:rho0} components can be rewritten in a similar way.

\section{Conclusion}

We have shown that in sufficiently strong magnetic field the empty space becomes an electromagnetic superconductor. The new state of the vacuum has many unusual features~\cite{Chernodub:2010qx,Chernodub:2011mc,ref:Jos,Chernodub:2012bj}: 

\begin{itemize}

\item The magnetic field induces the superconductivity instead of destroying it. 

\item The Meissner effect is absent. 

\item The superconductivity has a strong anisotropy: the electric currents may flow without resistance only along the axis of the magnetic field. 

\item The superconductivity appears in the empty space as a result of the restructuring of the quantum fluctuations due to the presence magnetic field. The overcritical magnetic field ($B > B_{c} \approx 10^{16}$\,T) induces the quark--antiquark condensates which have the quantum numbers of the electrically charged $\rho$-mesons.

\item The electromagnetic superconductivity is always accompanied by the superfluid component caused by emergence of a neutral $\rho$--meson condensate.

\item The tandem superconductor-superfluid ground state is inhomogeneous, it resembles an Abrikosov lattice in a mixed state of an ordinary type--II superconductor.

\item The charged and neutral vector quark-antiquark condensates have stringlike topological singularities: superconductor and superfluid $\rho$ vortices, respectively.

\item The ground state is characterized by a ``kaleidoscopic'' lattice structure made of the equilateral triangular lattice of the superconductor vortices which is superimposed on the hexagonal lattice of the superfluid vortices.

\end{itemize}

The vacuum superconductivity may be considered as a ``magnetic'' analogue of the Schwinger effect. Indeed, the Schwinger effect ({\emph {the vacuum superconductivity}}) is the electron-positron pair production ({\emph {the emergence of the quark-antiquark condensates}}) due to strong electric ({\emph {magnetic}}) field background in the vacuum. Contrary to the Schwinger effect, the vacuum superconductivity is a state, not a process.

The sufficiently strong magnetic fields, of the strength from two to three times higher than the required critical value~\eq{eq:Bc} may emerge in the ultraperipheral heavy-ion collisions at the Large Hadron Collider (LHC) at CERN~\cite{Deng:2012pc}. Thus, signatures of the magnetic-field-induced superconductivity have a chance to be found in laboratory conditions.

\end{document}